\RequirePackage{fix-cm}
\documentclass[smallextended]{svjour3}       % onecolumn (second format)

\smartqed

 \smartqed
 
\usepackage{graphicx}
\usepackage{amsmath}
\usepackage{amssymb}
\usepackage{booktabs}
\usepackage{multirow}
\usepackage{hyperref}
\usepackage{xcolor}
\usepackage{listings}
\usepackage{enumitem}
\usepackage{array}
\usepackage{tabularx}
\usepackage{float}
\usepackage[most]{tcolorbox}
\usepackage{tikz}
\usetikzlibrary{decorations.pathreplacing,calligraphy,positioning,arrows.meta,calc}
\usepackage{forest}
\usetikzlibrary{calc,decorations.pathreplacing,calligraphy}

\definecolor{promptbg}{HTML}{E8F4FC}
\definecolor{promptheader}{HTML}{1A7EBC}
\definecolor{reasonbg}{HTML}{FDE9D9}
\definecolor{reasonheader}{HTML}{F79646}
\definecolor{codebg}{HTML}{F3F0F6}
\definecolor{codeheader}{HTML}{8064A2}
 
\newtcolorbox{promptbox}{
  colback=promptbg,
  colframe=promptheader,
  coltitle=white,
  fonttitle=\bfseries\small,
  title=Prompt,
  boxrule=0.5pt,
  arc=1pt,
  left=4pt, right=4pt, top=2pt, bottom=2pt,
  before skip=4pt, after skip=4pt
}
 
\newtcolorbox{reasonbox}{
  colback=reasonbg,
  colframe=reasonheader,
  coltitle=white,
  fonttitle=\bfseries\small,
  title=Reasoning,
  boxrule=0.5pt,
  arc=1pt,
  left=4pt, right=4pt, top=2pt, bottom=2pt,
  before skip=4pt, after skip=4pt
}
 
\newtcolorbox{codebox}{
  colback=codebg,
  colframe=codeheader,
  coltitle=white,
  fonttitle=\bfseries\small,
  title=PlantUML Code,
  boxrule=0.5pt,
  arc=1pt,
  left=4pt, right=4pt, top=2pt, bottom=2pt,
  before skip=4pt, after skip=4pt
}
 
\newtcolorbox{echobox}{
  colback=reasonbg,
  colframe=reasonheader,
  boxrule=0.5pt,
  arc=1pt,
  left=4pt, right=4pt, top=4pt, bottom=4pt,
  before skip=4pt, after skip=4pt
}
 
\usepackage[dvipsnames]{xcolor}

\smartqed  % flush right qed marks, e.g. at end of proof
\usepackage{graphicx}
\usepackage{natbib}

\begin{document}

\title{Tracing Hallucinations in LLM Reasoning for UML Class Diagram Generation: An Empirical Taxonomy and Mitigation Study%\thanks{Grants or other notes
}
\subtitle{}

\titlerunning{Tracing Hallucinations in LLM Reasoning for UML Class Diagram Generation}     % if too long for running head

\author{ Mojdeh Rahmanian         \and
        Foutse Khomh \and
        Yanchao Yu
        \and
        Ashkan Sami 
}


\date{}

%%%%%For preprint
\makeatletter
\def\makeheadbox{}
\makeatother
%%%%%%

\maketitle

\begin{abstract}
Large Language Models (LLMs) can generate Unified Modeling Language (UML) class diagrams from natural-language requirements, but hallucinations may compromise the correctness and completeness of the resulting diagrams. Existing evaluations mostly assess the final diagram, without examining the intermediate traces of reasoning-capable LLMs, which can reveal where these hallucinations arise. We develop an empirical taxonomy of eleven hallucination sub-types organised into three dimensions: \emph{Prompt Faithfulness} (deviations between requirements and reasoning), \emph{Code Faithfulness} (discrepancies between reasoning and generated PlantUML code), and \emph{Reasoning Quality} (problems within the reasoning itself). The taxonomy was derived from 60 manually annotated DeepSeek-R1 traces, applied to 120 further DeepSeek-R1 traces through a dual-LLM annotation pipeline, and examined on 20 DeepSeek-V4-Pro traces. Code Faithfulness was the most frequent dimension for DeepSeek-R1 in both datasets, whereas Prompt Faithfulness dominated for DeepSeek-V4-Pro, suggesting model-dependent failure patterns. In the manually annotated traces, reasoning-trace length showed no significant monotonic association with hallucination frequency, whereas requirement complexity was negatively associated with it, particularly for Prompt Faithfulness errors. In a mitigation experiment on 13 Golden UML ModelSet cases with three Qwen3-based models, adding Prompt Faithfulness and Reasoning Quality constraints to the generation prompt raised the pooled mean $F_1$ score by 4.0 percentage points, whereas a post-hoc Code Faithfulness review provided no consistent further benefit. Relationship modelling remained the lowest-scoring element throughout. These results suggest that localising hallucinations within the generation process can inform targeted, prompt-based mitigation. Whether these findings generalise beyond the DeepSeek and Qwen3 model families remains to be tested.

\end{abstract}
 
%----------------------------------------------------------------------

\section{Introduction}
\label{sec:introduction}
Large language models (LLMs) have rapidly become capable tools for automating a wide range of software engineering activities, including requirements analysis, code generation, and software design~\cite{li2022competition,ataei2025elicitron,fan2023large,hou2024large}. A particularly promising application is the generation of Unified Modeling Language (UML) class diagrams from natural-language requirements~\cite{debari2024evaluating,camara2024systematic}. Such generation addresses a fundamental challenge in software engineering: translating informal, often incomplete textual requirements into explicit structural representations that can guide subsequent design and implementation. Reliable automation at this stage could accelerate early design activities, strengthen requirements traceability, and support both professional modeling and software engineering education.

The usefulness of such automation, however, depends not only on whether an LLM can produce a syntactically valid diagram, but on whether the diagram faithfully represents the requirements. Existing studies show that LLM-generated UML models can contain hallucinated classes, attributes, operations, or relationships that are unsupported by, inconsistent with, or absent from the requirements~\cite{ji2023survey,huang2025survey}. Recent empirical evaluations show that while LLM outputs are often syntactically valid and understandable, they suffer from significant issues related to completeness and correctness, particularly when requirements are ambiguous or complex~\cite{debari2024evaluating,ferrari2024model,camara2024systematic}.  This is particularly problematic for software models because structural errors can be deceptively plausible: a fabricated class, an unsupported association, or an incorrect multiplicity may pass superficial inspection while nevertheless influencing downstream design decisions, code generation, consistency checking, and design review. Understanding and preventing these errors is therefore essential if LLM-based modeling is to move from demonstration to dependable software engineering practice.

Hallucination in UML generation also differs fundamentally from hallucination in conventional text generation. Research has extensively studied hallucinations in tasks such as summarisation~\cite{maynez2020faithfulness}, machine translation~\cite{raunak2021curious}, and code synthesis~\cite{zhang2025llm}. In these settings, hallucination is commonly framed as a mismatch between generated content and a reference source or specification. UML generation introduces an additional challenge: the output is a structured artefact whose correctness depends on multiple transformations between the input requirements, the model's internal reasoning, and the representation used to encode the diagram. Errors may therefore arise not only in the final artefact but at each of these transformations, and may propagate from one to the next.

This process-level view is particularly relevant for reasoning-capable LLMs. Rather than mapping requirements directly to a diagram, these models often generate explicit reasoning traces in which they identify candidate entities, infer attributes and operations, establish relationships, determine multiplicities, and finally translate these decisions into diagram code. This process creates several distinct failure points: an error introduced when interpreting the requirements can propagate through subsequent reasoning; individually plausible reasoning steps can collectively produce an unsound model; and a sound reasoning process can be corrupted during translation into PlantUML. Yet existing evaluations of LLM-generated UML diagrams largely focus on the quality of the final artefact, providing limited insight into where in the reasoning-to-diagram process hallucinations arise, how they propagate, and which failure modes are most prevalent. This lack of process-level understanding limits our ability to design targeted mitigation strategies. For example, although prompt engineering~\cite{freeman2026epistemic,zhang2025mitigating}, retrieval-augmented generation~\cite{lewis2020retrieval,shuster2021retrieval}, and post-hoc verification~\cite{dhuliawala2024chain,manakul2023selfcheckgpt} may address different failure mechanisms, their effectiveness depends on knowing which mechanisms actually drive UML generation errors. A systematic characterisation of these failure modes is therefore a necessary first step toward both understanding and improving the reliability of LLM-generated UML models.

In this paper, we address this gap by constructing and validating an empirical taxonomy of hallucinations in reasoning-capable LLMs generating UML class diagrams, and by demonstrating how the taxonomy can be translated into actionable mitigation strategies. The taxonomy identifies three dimensions corresponding to the generation process: \emph{Prompt Faithfulness}, capturing deviations between the requirements and the model's reasoning; \emph{Reasoning Quality}, capturing unsoundness within the reasoning chain; and \emph{Code Faithfulness}, capturing discrepancies between the reasoning and the resulting PlantUML representation. Across these dimensions, we identify eleven hallucination sub-types. We construct the taxonomy through manual analysis of reasoning traces, examine its distribution and generalisability using an independent dataset and a dual-LLM annotation pipeline, investigate the relationship between hallucination frequency and input and reasoning characteristics, and operationalise the taxonomy as generation-time constraints and post-hoc verification mechanisms across multiple reasoning-capable models.

To structure our investigation, we address four research questions:
\begin{description}
\item[RQ1.] What categories of hallucinations arise during the reasoning traces of LLMs generating UML class diagrams, and how can they be organised into a taxonomy?
\textit{We identify three dimensions, namely Prompt Faithfulness, Code Faithfulness, and Reasoning Quality, with eleven sub-types corresponding to distinct failure modes in the requirements-to-reasoning-to-code pipeline.}

\item[RQ2.] How are hallucinations distributed across the taxonomy categories, and is the distribution stable when the taxonomy is applied to an independent dataset through automated annotation?
\textit{Code Faithfulness is the most frequent dimension in both DeepSeek-R1 datasets, and the dimension-level ordering is consistent across annotation regimes.}

\item[RQ3.] How are hallucination frequencies related to the length of the model's reasoning trace and the complexity of the input requirements?
\textit{Reasoning-trace length shows no significant monotonic association with hallucination frequency, while requirement complexity is negatively associated with overall hallucination frequency and Prompt Faithfulness errors.}

\item[RQ4.] To what extent can embedding taxonomy-derived rules into the generation and verification pipeline improve the quality of LLM-generated UML class diagrams?
\textit{Taxonomy-derived constraints embedded in the generation prompt improve mean diagram quality across all evaluated models, while an additional post-hoc Code Faithfulness review provides no consistent additional benefit.}
\end{description}

The paper makes four contributions:

\begin{enumerate}
  \item An evidence-based, empirical taxonomy of hallucination types in UML class-diagram generation, organised along three dimensions that reflect the multi-stage reasoning process.

  \item A dual-LLM evaluation pipeline for automated hallucination detection and verification at scale, without requiring human ground-truth annotations for each evaluated trace, with its consistency assessed against human annotations on a sample of traces.

  \item A quantitative analysis of hallucination distributions and their relationships with reasoning-trace length and requirement complexity.

  \item An empirical evaluation of a taxonomy-guided generation and verification pipeline across three reasoning-capable models, showing improvements in mean diagram quality from taxonomy-aware generation, while finding no consistent additional benefit from post-hoc Code Faithfulness review.
\end{enumerate}

The remainder of this paper is organised as follows. Section~\ref{sec:background} reviews related work on hallucination and LLMs in software engineering. Section~\ref{sec:design} presents the study design for RQ1--RQ3, including the datasets, annotation procedures, and analyses. Section~\ref{sec:results} reports the results for RQ1--RQ3; it first presents the hallucination taxonomy and its definitions with illustrative examples (Section~\ref{sec:HallucinationTaxonomy}), and then reports the distribution of hallucinations and their associations with reasoning length and requirement complexity. Section~\ref{sec:TaxMitigation} describes the design and results of the taxonomy-guided mitigation experiment addressing RQ4. Section~\ref{sec:discussion} synthesises the findings and discusses their practical implications. Finally, Sections~\ref{sec:threats} and~\ref{sec:conclusion} discuss threats to validity and conclude the paper.
%---------------------------------------------------

\section{Background \& Related Work}
\label{sec:background}

\subsection{Hallucination in Large Language Models}

In natural language generation (NLG), hallucination refers to the production of fluent output that does not appear in source materials or is factually incorrect. Comprehensive surveys compile causes across tasks such as summarisation, question answering, machine translation, and dialogue, distinguishing intrinsic hallucinations (conflicting with the source) from extrinsic ones (non-verifiable with the source)~\cite{ji2023survey}.
%%%again
Recent LLM-focused overviews propose refined taxonomies, benchmark families, and mitigation strategies such as retrieval grounding, self-consistency, and uncertainty estimation, emphasising reliability challenges unique to general-purpose LLMs~\cite{huang2025survey}.

Many early investigations focused on purely textual tasks~\cite{raunak2021curious,maynez2020faithfulness}. However, more recent work has identified hallucinations in structured tasks as well, including code generation and symbolic reasoning~\cite{zhang2025llm}. In these contexts, hallucinations may be syntactically correct but semantically incorrect, for example, fabricated variable dependencies or invented class hierarchies. This distinction is particularly relevant to model-based generation, where the output must satisfy both notational syntax and domain semantics.

\subsection{Taxonomies of LLM Errors and Hallucinations}

The construction of empirical taxonomies has a long tradition in software-engineering research, beginning with Orthogonal Defect Classification~\citep{chillarege1992orthogonal} and extending to more recent classifications of faults in machine-learning systems~\citep{humbatova2020taxonomy}. Such taxonomies serve as analytical instruments that enable systematic diagnosis, benchmark construction, and the design of targeted interventions, in contrast to single-metric evaluations that report \emph{whether} a system failed without characterising \emph{how}.

The same methodology has been adopted to characterise the failure modes of LLMs. For code generation, \citet{liu2024exploring} examined hallucinations in code produced by several LLMs and established a taxonomy of five primary categories organised around the sources of semantic conflict (intent, contextual code, and world knowledge). \citet{pan2024lost} catalogued 15 categories of bugs that LLMs introduce when translating between programming languages, identified through manual analysis of 1{,}748 labelled defects across five languages. For natural-language tasks, multi-dimensional schemes covering factuality, faithfulness, and consistency have been developed for dialogue, summarisation, and retrieval-augmented generation, building on the intrinsic-versus-extrinsic distinction introduced by \citet{maynez2020faithfulness} and later synthesised in surveys by \citet{ji2023survey} and \citet{huang2025survey}.

Beyond output faithfulness, a growing body of work examines the reasoning traces themselves: whether they faithfully reflect the model's underlying computation and whether they are internally sound. \citet{turpin2023language} showed that chain-of-thought (CoT) explanations can systematically misrepresent the true reason for a model's prediction, for example when the answer is influenced by biasing features of the prompt that the explanation never mentions. \citet{lanham2023measuring} introduced perturbation-based methods to measure this faithfulness gap in CoT reasoning. More recently, attention has shifted to large reasoning models (LRMs), which produce extended thinking traces before generating a final answer. \citet{marjanovic2025deepseek} conducted an in-depth analysis of DeepSeek-R1 reasoning chains and documented recurring behaviours, including speculative branching, self-contradiction, and repeated reversal of decisions, which closely resemble sub-types in our Reasoning Quality dimension. \citet{lee_evaluating_2025} surveyed methods for evaluating step-by-step reasoning traces and organised the evaluation criteria into groundedness, validity, coherence, and utility. \citet{sun_detection_2025} investigated hallucination in LRMs from a mechanistic perspective, showing that hallucinations arising in reasoning traces can have causes distinct from those observed in conventional LLM outputs. Their findings highlight the importance of examining hallucinations at the reasoning-process level rather than treating them solely as properties of the final output.

For structured-generation tasks that map natural-language requirements to symbolic artefacts, however, this perspective remains largely unexplored. As discussed in the next subsection, existing studies of LLM-based UML generation report hallucinations and structural errors but primarily evaluate them at the output level. To the best of our knowledge, no prior work systematically characterises these failures as a taxonomy derived from the reasoning traces produced during UML generation.

\subsection{LLMs for UML and Software Modelling}

LLMs are increasingly applied across software engineering workflows, from requirements analysis to code generation and design~\citep{fan2023large,hou2024large}. In code generation, AlphaCode achieved approximately median-level performance in programming competitions, highlighting both the promise and the limits of LLM-driven program synthesis~\citep{li2022competition}. Within requirements and design activities, recent studies suggest that LLMs can support early modelling tasks by interpreting natural-language requirements and generating design representations~\citep{ataei2025elicitron,belzner2023large}.

The specific task of translating natural-language descriptions into UML diagrams has received growing attention. \citet{camara2024systematic} reported on their experience of using ChatGPT for UML class-diagram generation and found strong performance on OCL constraints but critical limitations in scalability (beyond 10--12 classes), poor support for association classes, and high variability across prompts. \citet{ferrari2024model} examined sequence-diagram generation from 28 real-world requirements documents, reporting reasonable conformance to UML standards but significant degradation in completeness and correctness when requirements are ambiguous. \citet{debari2024evaluating} evaluated LLMs on 20 class-diagram modelling exercises and found that, while outputs were generally understandable, issues of completeness and correctness remained, particularly for complex exercises. \citet{conrardy2024image} explored image-based UML generation and observed that even vision-enabled LLMs introduce structural errors when interpreting diagram images.

A common finding across these studies is that LLM-generated UML diagrams frequently contain unsupported, missing, or contradictory elements, which we treat as hallucinations in this study. However, these works report such errors as observed phenomena at the output level, without analysing their types or origins. As noted above, no prior work offers a systematic taxonomy that explains \emph{how} these hallucinations arise during the multi-stage reasoning process that reasoning-capable LLMs follow when generating diagrams.

\subsection{Hallucination Mitigation}

Existing hallucination mitigation strategies can be broadly grouped into four families~\citep{huang2025survey}: (i)~training-time interventions, such as reinforcement learning from human feedback (RLHF) and direct preference optimisation (DPO); (ii)~inference-time augmentation, such as retrieval-augmented generation (RAG)~\citep{lewis2020retrieval,shuster2021retrieval}; (iii)~post-hoc verification and self-correction, in which the model checks and revises its own output after generation~\citep{dhuliawala2024chain,manakul2023selfcheckgpt}; and (iv)~prompt engineering, which reshapes the input to a fixed model without modifying its weights.

Prompt engineering is arguably the most accessible approach for practitioners using hosted, black-box models, and recent work has explored structured prompting to reduce hallucinations in various domains. \citet{zhang2025mitigating} demonstrated that decomposing spatial reasoning tasks into structured sub-steps significantly reduced hallucinations in path planning. \citet{freeman2026epistemic} compared five prompt engineering strategies for industrial LLM applications and found that structured procedures with explicit verification steps produced more consistent, grounded outputs. \citet{shim2025cpr} proposed Curative Prompt Refinement, which reformulates ill-formed prompts to reduce hallucinatory outputs.

In the software modelling domain, however, mitigation efforts remain limited. Prior UML generation studies~\citep{debari2024evaluating,camara2024systematic,ferrari2024model} have reported structural errors but have not systematically attempted to prevent them. Where prompting strategies have been explored for modelling tasks, they have relied on generic techniques such as few-shot or chain-of-thought prompting without domain-specific error awareness~\citep{chen2023automated}. This gap motivates the mitigation experiment addressing RQ4 (Section~\ref{sec:TaxMitigation}), which operationalises the hallucination taxonomy as generation-time constraints and post-hoc verification rules, and evaluates whether these taxonomy-derived interventions improve diagram quality.
%-------------------------------
\section{Study Design}
\label{sec:design}
 
This section describes the datasets, procedures, and evaluation methods used to address each research question. Three benchmark datasets are used across the studies, selected to serve complementary purposes.
 
\paragraph{Dataset~1: ~\cite{debari2024evaluating}.} This dataset consists of 20 modelling exercises drawn from open-source educational materials. Each exercise provides a natural-language description and a validated reference UML class diagram. The exercises were selected based on two criteria: (i)~the description was in a collector-readable language (English or Italian), and (ii)~an accepted reference solution existed. Table~\ref{tab:exercises} summarises the structural characteristics of each exercise. This dataset is used for manual taxonomy construction (RQ1) and hallucination distribution analysis (RQ2).
 
\paragraph{Dataset~2: ~\cite{bozyigit2024generating}.} This dataset contains 120 domain-modelling descriptions from multiple domains (e-commerce, healthcare, education, logistics, and others), with broader lexical and structural diversity than Dataset~1. It is used to assess the generalisability and scalability of the taxonomy through automated dual-LLM evaluation (RQ2).
 
\paragraph{Dataset~3: Golden UML ModelSet~\cite{verbruggen2025golden}.} This is an expert-curated collection of UML class diagrams with verified PlantUML gold-standard solutions. We selected 13 cases, one from each domain in the dataset: Business Services, Education, Financial Services, Healthcare, Insurance, Leisure and Recreation, Logistics, Manufacturing, Media and Publishing, Personal Activities, Research, Sales, and Social Networks. This dataset is used exclusively for the mitigation experiment (RQ4), where gold solutions enable objective evaluation of diagram quality. Table~\ref{tab:gold_cases} summarises the structural characteristics of the selected cases.
 
\begin{table}[t]
\caption{Structural characteristics of Dataset~1 exercises~\cite{debari2024evaluating}.}
\label{tab:exercises}
\centering
\small
\begin{tabular}{lccc}
\hline\noalign{\smallskip}
Exercise & Classes & Attr+Ops & Associations \\
\noalign{\smallskip}\hline\noalign{\smallskip}
1  & 6  & 17 & 7  \\
2  & 6  & 35 & 5  \\
3  & 7  & 12 & 8  \\
4  & 6  & 31 & 7  \\
5  & 8  & 18 & 11 \\
6  & 9  & 12 & 12 \\
7  & 6  & 5  & 7  \\
8  & 7  & 11 & 8  \\
9  & 9  & 13 & 10 \\
10 & 7  & 15 & 7  \\
11 & 6  & 11 & 9  \\
12 & 6  & 11 & 5  \\
13 & 8  & 19 & 8  \\
14 & 5  & 12 & 5  \\
15 & 8  & 6  & 8  \\
16 & 11 & 13 & 13 \\
17 & 8  & 5  & 8  \\
18 & 6  & 10 & 6  \\
19 & 9  & 20 & 9  \\
20 & 10 & 28 & 10 \\
\noalign{\smallskip}\hline
\end{tabular}
\end{table}

\begin{table}[t]
\caption{Structural characteristics of the 13 Golden UML ModelSet cases, ordered by complexity (Classes + Associations).}
\label{tab:gold_cases}
\centering
\small
\begin{tabular}{lcc}
\hline\noalign{\smallskip}
Case & Classes & Associations \\
\noalign{\smallskip}\hline\noalign{\smallskip}
BusTransportation         & 5  & 4  \\
BankAccount               & 6  & 5  \\
Sightseeing               & 7  & 7  \\
ClothingCompany           & 7  & 9  \\
FitnessCompanyConan       & 8  & 9  \\
University                & 9  & 9  \\
EUScienceConnect          & 8  & 10 \\
eHome2020                 & 8  & 10 \\
AlphaInsurance            & 10 & 10 \\
Facepage                  & 9  & 11 \\
BuildingManagement        & 10 & 11 \\
LabTracker                & 12 & 12 \\
TileOGame                 & 16 & 21 \\
\noalign{\smallskip}\hline
\end{tabular}
\end{table}
 
\subsection{RQ1: Taxonomy Construction}
\label{sec:design_study1}
 
To answer RQ1, we constructed an empirical taxonomy of hallucinations in UML class-diagram generation through systematic analysis of model reasoning traces and outputs.
 
\paragraph{Model and prompting.} Model queries were issued to DeepSeek-R1, a reasoning-capable LLM that produces explicit chain-of-thought traces. Each natural-language exercise description from Dataset~1 was submitted with a standardised instruction:
 
\begin{quote}
\texttt{Create a UML Class Diagram of the given exercise and give me the PlantUML code.\\
Input: exercise: \{Desc\}}
\end{quote}
 
Each exercise was executed three times under different decoding configurations corresponding to low, medium, and high stochasticity, implemented by adjusting the temperature parameter (0, 0.6, and 1.0, respectively). In total, 60 model runs were obtained (20~exercises $\times$ 3~temperature settings).
 
\paragraph{Artefacts.} Each run yielded two paired artefacts: (1)~the model's internal reasoning trace, delimited as \texttt{<think> \ldots\ </think>}, and (2)~the final PlantUML class-diagram code block (delimited by \texttt{@startuml \ldots\ @enduml}).
 
\paragraph{Coding procedure.}
In a preliminary step, two authors jointly examined 20 reasoning traces, together with their corresponding prompts and final diagrams, through qualitative open
coding. Whenever the model introduced unsupported details, contradicted its earlier statements, or produced invalid UML constructs, the phenomenon was noted. Initial category labels and operational definitions were developed, and overlapping labels were merged through discussion to
establish a preliminary codebook. The remaining 40 traces were then independently annotated by one author and a
master's student using the codebook. Disagreements were resolved through discussion to reach consensus.
 
\subsection{RQ2 \& RQ3: Taxonomy Validation and Predictor Analysis}
\label{sec:design_study2}
 
To answer RQ2, we validated the taxonomy's generalisability across a larger, independent dataset through automated evaluation. To answer RQ3, we analysed how hallucination patterns relate to reasoning-trace length and requirement complexity.
 
\paragraph{Dual-LLM evaluation.} To enable scalable hallucination detection without reliance on human ground truth, the manually established taxonomy was operationalised through a dual-LLM evaluation system. The first model (LLM as Annotator) identified suspected hallucinations by applying the taxonomy categories to each reasoning trace and PlantUML output. The second model (LLM as Judge) independently reviewed these annotations, confirming or rejecting each one and producing a verified set of hallucination counts per category. To ensure diversity of judgment and reduce model-specific bias, the annotator and judge were instantiated on different platforms: Claude~3.7~Sonnet (via Anthropic API) as annotator and GPT-4o (through OpenAI API) as judge. This approach builds on recent evidence that LLMs can serve as reliable evaluators with human-like agreement for open-ended assessments~\cite{zheng2023judging} and high consistency in code and text evaluation pipelines~\cite{tong2024codejudge,gu2025survey}. 

 %again
\paragraph{Data and scope.} The dual-LLM pipeline was applied 
to Dataset~2 (120 descriptions), enabling validation of the 
manually constructed taxonomy on a larger, independent corpus. To assess the consistency of the dual-LLM pipeline
with human judgement, one author independently annotated
a stratified random sample of 30 reasoning traces
(25\% of Dataset~2). The human annotations were compared
with the pipeline outputs using pairwise~$F_1$, matching
hallucination instances identified at the same location
in the trace.
 
\paragraph{Correlation analyses.} To address RQ3, we computed two sets of correlations. First, the relationship between reasoning-trace length (in tokens) and the average number of hallucination errors across the three taxonomy dimensions. Second, the relationship between requirement complexity, computed using the formula proposed by ~\cite{bozyigit2024generating}, and hallucination frequency. The complexity score combines four textual indicators: the number of sentences (\textit{NoS}), total words (\textit{NoW}), words after removing stop words (\textit{NoSW}), and the number of unique words (\textit{NoUW}):
\begin{equation}
\label{eq:complexity}
\text{Complexity} = \frac{(0.4 \times NoS) + (0.2 \times NoW) + (0.2 \times NoUW) + (0.2 \times NoSW)}{10}
\end{equation}

\section{Results}
\label{sec:results}

This section presents the findings for RQ1, RQ2, and RQ3.

\subsection{RQ1: A Taxonomy of Hallucinations in LLM-Generated UML Class Diagrams}
\label{sec:taxonomy}

Identifying hallucinations in the reasoning traces of UML class-diagram generation is inherently difficult. DeepSeek-R1 reasoning traces are long and discursive, averaging approximately 4{,}800 tokens per trace (range 506--9{,}632). Identifying a hallucination requires the annotator to simultaneously hold the original prompt, the evolving reasoning chain, and the final PlantUML output in mind, and to judge whether a deviation is a genuine unsupported inference or a legitimate elaboration.

To address this difficulty, the taxonomy was developed through a two-stage coding procedure (Section~\ref{sec:design_study1}). In the first stage, two authors jointly examined 20 reasoning traces through open coding to elicit an initial set of category labels and operational definitions; overlapping labels were merged, and a stable codebook was produced through discussion. During this stage, initial observations suggested the presence of circular reasoning patterns; however, this candidate sub-type was ultimately excluded from the codebook because the raters could not agree on its operational boundaries. In the second stage, the codebook was applied independently to the remaining 40 traces by the first author and a master's student in computer science, and disagreements were resolved through discussion until consensus was reached. The complete set of reasoning traces, generated PlantUML code, and annotations is available in our replication package.\footnote{\url{https://github.com/MojdehRahmanian/HallucinationTaxonomy}}

The resulting taxonomy organises hallucinations along three dimensions, each mapped to a distinct transition in the requirements-to-reasoning-to-code process (Figure~\ref{fig:taxonomy}). The first dimension, \emph{Prompt Faithfulness}, captures hallucinations at the prompt-to-reasoning transition and contains four sub-types: Unsupported Entity, Unsupported Relation, Over-specific Detail, and Missed Requirement. The second dimension, \emph{Code Faithfulness}, captures hallucinations at the reasoning-to-code transition and contains four sub-types: Dropped Step, Fabricated Code, Contradiction, and Structural Mismatch. The third dimension, \emph{Reasoning Quality}, captures hallucinations internal to the reasoning chain and contains three sub-types: Speculative Branch, Goal Drift, and Tool Misuse.

\begin{figure}[t]
\centering
\resizebox{\linewidth}{!}{%
\begin{tikzpicture}[
  font=\footnotesize,
  leaf/.style={
    draw, rounded corners=2pt,
    text width=3.6cm,
    minimum height=0.55cm,
    align=center, anchor=west,
    inner xsep=5pt, inner ysep=3pt,
  },
  cat/.style={
    draw, rounded corners=2pt, fill=gray!12,
    minimum width=2.6cm, minimum height=0.95cm,
    align=center, font=\footnotesize\bfseries,
  },
  root/.style={
    draw, rounded corners=2pt, fill=black!85, text=white,
    minimum width=0.75cm, minimum height=4.2cm,
    inner sep=2pt,
  },
  myedge/.style={thin, draw=black!75},
]

\node[root] (root) at (0,0)
  {\rotatebox{90}{\bfseries Class Diagram Generation Hallucination}};

\node[cat] (cat1) at (3.0,  3.00) {Prompt\\Faithfulness};
\node[cat] (cat2) at (3.0,  0.00) {Code\\Faithfulness};
\node[cat] (cat3) at (3.0, -2.70) {Reasoning\\Quality};

\node[leaf] (a1) at (6.2,  3.90) {A1: Unsupported Entity};
\node[leaf] (a2) at (6.2,  3.30) {A2: Unsupported Relation};
\node[leaf] (a3) at (6.2,  2.70) {A3: Over-specific Detail};
\node[leaf] (a4) at (6.2,  2.10) {A4: Missed Requirement};

\node[leaf] (b1) at (6.2,  0.90) {B1: Dropped Step};
\node[leaf] (b2) at (6.2,  0.30) {B2: Fabricated Code};
\node[leaf] (b3) at (6.2, -0.30) {B3: Contradiction};
\node[leaf] (b4) at (6.2, -0.90) {B4: Structural Mismatch};

\node[leaf] (c1) at (6.2, -2.10) {C1: Speculative Branch};
\node[leaf] (c2) at (6.2, -2.70) {C2: Goal Drift};
\node[leaf] (c3) at (6.2, -3.30) {C3: Tool Misuse};

%% ---------- Root -> Categories ----------
\draw[myedge] (root.east) -- (cat1.west);
\draw[myedge] (root.east) -- (cat2.west);
\draw[myedge] (root.east) -- (cat3.west);

%% ---------- Categories -> Leaves ----------
\foreach \cat/\top/\bot/\list in {%
  cat1/a1/a4/{a1,a2,a3,a4},
  cat2/b1/b4/{b1,b2,b3,b4},
  cat3/c1/c3/{c1,c2,c3}%
}{
  \coordinate (j) at ($(\cat.east) + (0.35,0)$);
  \draw[myedge] (\cat.east) -- (j);
  \draw[myedge] (j |- \top) -- (j |- \bot);
  \foreach \n in \list {
    \draw[myedge] (j |- \n) -- (\n.west);
  }
}

\def\braceX{10.4}

\draw[decorate, decoration={calligraphic brace, amplitude=5pt, raise=3pt}, thick]
  (\braceX,  4.175) -- (\braceX,  1.825)
  node[midway, right=11pt, font=\footnotesize\itshape]
  {Prompt $\rightarrow$ Reasoning};

\draw[decorate, decoration={calligraphic brace, amplitude=5pt, raise=3pt}, thick]
  (\braceX,  1.175) -- (\braceX, -1.175)
  node[midway, right=11pt, font=\footnotesize\itshape]
  {Reasoning $\rightarrow$ Diagram};

\draw[decorate, decoration={calligraphic brace, amplitude=5pt, raise=3pt}, thick]
  (\braceX, -1.825) -- (\braceX, -3.575)
  node[midway, right=11pt, font=\footnotesize\itshape]
  {Within Reasoning};

\end{tikzpicture}%
}
\caption{Taxonomy of hallucinations in LLM-generated UML class diagrams, organised along the three transitions of the generation pipeline.}
\label{fig:taxonomy}
\end{figure}

To assess pre-consensus inter-rater consistency on the 40 independently annotated traces, we calculated pairwise~$F_1$~\cite{hripcsak2005agreement} based on the hallucination instances identified by the two annotators. An instance was considered matched when both annotators identified the same hallucination at the same location in the trace. Here, $n$ denotes the number of distinct instances identified by at least one annotator within each dimension, and $N$ denotes the total across all three dimensions. Pairwise~$F_1$ was 79.78\% for Prompt Faithfulness ($n=107$), 94.04\% for Code Faithfulness ($n=160$), and 61.90\% for Reasoning Quality ($n=116$), with an overall micro-averaged~$F_1$ of 81.79\% ($N=383$). The lower consistency for Reasoning Quality reflects the difficulty noted above: judging whether a reasoning step is unsound requires assessing the reasoning process itself rather than comparing it with the prompt or the generated code. The first 20 traces were coded jointly during codebook development and therefore have no independent pre-consensus annotations; all subsequent analyses (RQ2 and RQ3) use the consensus annotations of all 60 traces.

Section~\ref{sec:HallucinationTaxonomy} defines each sub-type and illustrates it with a representative example from the coded traces.

\subsection{Hallucination Taxonomy}\label{sec:HallucinationTaxonomy}

The following subsections define the sub-types of each dimension, specifying the unit of analysis and the decision rule applied during annotation, and illustrate each sub-type with a representative example from the coded traces.

\subsubsection{Dimension A: Prompt Faithfulness (Prompt $\to$ Reasoning)}
 
This dimension captures cases where the model's reasoning chain introduces or omits content relative to the input requirements, independent of the final PlantUML output. We apply an evidence rule: every committed element in the chain (entity, relation, attribute, multiplicity, operation) must be justified by an explicit span in the prompt.
 
\paragraph{A1, Unsupported Entity.}We label an Unsupported Entity when the chain introduces a class or interface that has no explicit textual basis in the exercise description. The unit of analysis is a single classifier (class/interface).

Figure~\ref{fig:ex:a1} illustrates this pattern. The model introduces the \texttt{Supplier} and \texttt{RestockOrder} classes, even though the prompt does not mention these concepts. The prompt only describes operations related to customers, subscriptions, and movie ordering, not supply-chain interactions. These additional entities are therefore considered unsupported, illustrating how the model fabricates new classes that extend beyond the given requirements.
 
\begin{figure}[!ht]
\begin{promptbox}
\small Design a system for a movie-shop, in order to handle ordering of movies and browsing of the catalogue of the store, and user subscriptions with rechargeable cards. Only subscribers are allowed hiring movies with their own card. Credit is updated on the card during rent operations. Both users and subscribers can buy a movie and their data are saved in the related order. When a movie is not available it is ordered.
\end{promptbox}
 
\begin{reasonbox}
\small But note: the ordering here might be different: we need to handle reordering the movie from a supplier? So we might need:
\begin{itemize}[nosep]
  \item Supplier: to order movies when they are not available.
  \item RestockOrder: an order to a supplier to restock a movie.
\end{itemize}
\end{reasonbox}
\caption{A1 Unsupported Entity. The reasoning trace introduces \texttt{Supplier} and \texttt{RestockOrder} classes that have no textual basis in the requirement.}
\label{fig:ex:a1}
\end{figure}
 
  \paragraph{A2, Unsupported Relation.} We mark an Unsupported Relation when the reasoning trace asserts an association, generalisation, aggregation/composition, or a specific direction/multiplicity between entities without a justifying span in the requirement description. The unit of  analysis is a single relation. Implicit co-occurrence (``orders include items'') does not license composition, 
containment, or arrow direction unless the text states it; similarly, inheritance requires an explicit subtype cue.

Figure~\ref{fig:ex:a2} illustrates this. The model infers and commits to two associations (Project--Requirement and Project--System) with multiplicities and directionality, even though the requirement description only mentions that a project uses requirements and produces a system. The text does not specify ownership, creation, or multiplicity relations, yet the model interprets these as composition and assigns cardinalities. These added structural commitments are not supported by the requirement description and therefore constitute Unsupported Relations.

\begin{figure}[!ht]
\begin{promptbox}
\small \ldots As input, a project uses requirements. As output, a project produces a system (or part of a system). The requirements and system are work products: things that are created, used, updated, and elaborated on throughout a project. \ldots
\end{promptbox}
 
\begin{reasonbox}
\small But note: the project produces the system, so the system is created by the project. \ldots So let's assume the project creates the requirements as well? Then:
 
\texttt{Project "1" *-{}- "*" Requirement : requirements}
 
\texttt{Project "1" *-{}- "1" System : system}
\end{reasonbox}
\caption{A2 Unsupported Relation. The reasoning trace commits to composition relations with explicit multiplicities between Project and Requirement/System that the requirement description does not warrant.}
\label{fig:ex:a2}
\end{figure}
 
\paragraph{A3, Over-specific Detail.} We count Over-specific Detail when the reasoning trace invents multiplicities, datatypes, attributes, or operation signatures that go beyond what the requirement description specifies. The unit of analysis is a property, multiplicity, or signature. Numeric 
or cardinality cues in the text (e.g., ``exactly one'') justify commitments; otherwise, defaults like \texttt{1..*}, 
concrete types such as \texttt{DateTime}, or extra parameters are considered invented. Neutral placeholders (e.g., 
\texttt{\{type TBD\}}) are acceptable and not penalised.

Figure~\ref{fig:ex:a3} illustrates this. The model introduces additional details that are not supported by the requirement description, including specific attributes (\texttt{dateOfBirth}, \texttt{gender}), datatypes (\texttt{String}, \texttt{Date}), \\and an operation (\texttt{scheduleAppointment}) with parameters. The requirement description only mentions the presence of a patient record system but provides no attribute-level or procedural details. These added elements therefore represent Over-specific Details, meaning plausible yet fabricated information that extends beyond the given requirements.
 
\begin{figure}[!ht]
\begin{promptbox}
\small \ldots A patient record and scheduling system in a doctor's office is used by the receptionists, nurses, and doctors. The receptionists use the system to enter new patient information when first-time patients visit the doctor. \ldots
\end{promptbox}
 
\begin{reasonbox}
\small \ldots Key Entities and Their Attributes/Methods:
1. Patient:
   - Attributes: patientID, firstName, lastName, dateOfBirth, etc.
   - Relationships: belongs to a Family, has multiple Visits.
   
 \ldots
\begin{lstlisting}
class Patient {
  - patientID: String
  - firstName: String
  - dateOfBirth: Date
  - gender: String
  + scheduleAppointment(date: Date,
      time: Time, doctor: Doctor): Visit
}
\end{lstlisting}
\end{reasonbox}
\caption{A3 Over-specific Detail. The requirement refers only to ``new patient information'' without specifying what this information comprises. The reasoning trace nevertheless commits to concrete attributes for \texttt{Patient} (e.g., \texttt{dateOfBirth}, \texttt{gender}) and assigns them specific datatypes (\texttt{String}, \texttt{Date}), none of which is supported by the text.}
\label{fig:ex:a3}
\end{figure}

\paragraph{A4, Missed Requirement.} An element explicitly required by the prompt (entity, attribute, relation, or constraint) does not appear anywhere in the chain.

Figure~\ref{fig:ex:a4} illustrates this. The prompt explicitly states that tables are composed of rows \emph{and columns}, yet the model's reasoning includes only the classes Table, Row, and Cell. The absence of a \texttt{Column} class, which is directly implied by the prompt, indicates that the model omitted an element required by the specification.

\begin{figure}[!ht]
\begin{promptbox}
\small \ldots Tables are made up of rows and columns and every cell in a table can contain both text and pictures. \ldots
\end{promptbox}
 
\begin{reasonbox}
\small \ldots Now, for Table:
\begin{lstlisting}
class Table {
  rows: List<Row>
}
class Row {
  cells: List<Cell>
}
\end{lstlisting}
\end{reasonbox}
\caption{A4 Missed Requirement. Although the prompt mentions both rows and columns, the reasoning trace omits a \texttt{Column} class.}
\label{fig:ex:a4}
\end{figure}

\subsubsection{Dimension B: Code Faithfulness (Reasoning $\to$ PlantUML)}
 
This dimension evaluates agreement between the model's internal plan (the reasoning trace) and the final PlantUML output. We parse both into comparable fact sets of entities, attributes, operations, and relations.
 
\paragraph{B1, Dropped Step.} A planned element present in the reasoning trace is missing in the PlantUML code. The unit is the element at the finest granularity (class, attribute, operation, or relation). We mark ``dropped'' only when there is no equivalent implementation.
 
Figure~\ref{fig:ex:b1} illustrates this. The reasoning trace defines the class \texttt{Account} with two attributes, \texttt{accountNumber} and \texttt{balance}, but the corresponding PlantUML code includes only \texttt{balance}. The attribute \texttt{accountNumber}, which was planned in the reasoning trace, is absent from the final implementation.
 
\begin{figure}[!ht]
\begin{reasonbox}
\small
\begin{lstlisting}
...
class Account {
  - accountNumber: String
  - balance: double
}
abstract Account
...
\end{lstlisting}
\end{reasonbox}
 
\begin{codebox}
\small
\begin{lstlisting}
...
abstract class Account {
  - balance : Double
}
...
\end{lstlisting}
\end{codebox}
\caption{B1 Dropped Step. The reasoning trace plans both \texttt{accountNumber} and \texttt{balance} on \texttt{Account}, but the PlantUML code emits only \texttt{balance}.}
\label{fig:ex:b1}
\end{figure}

\paragraph{B2, Fabricated Code.} The PlantUML code introduces an element that the reasoning trace never committed to, including extra classes, relations, or attributes, as well as notes that carry modelling meaning. Minor syntactic sugar (e.g., whitespace, ordering) is ignored, but any semantic addition counts.
 
Figure~\ref{fig:ex:b2b3} illustrates B2 (and B3 below). The code introduces \texttt{Factory "1" -{}- "*" Machine} and \texttt{Machine "*" -{}- "*" Worker : operates}, neither of which appears in the reasoning plan.
 
\begin{figure}[!ht]
\begin{reasonbox}
\small
\begin{lstlisting}
...
Machine "1" --> "*" ProductType : produces
Client "1" --> "*" PurchaseOrder : places
PurchaseOrder "1" --> "*" OrderLine : contains
OrderLine "*" --> "1" ProductType : product
Worker "*" --> "*" Machine : operates
...
\end{lstlisting}
\end{reasonbox}
 
\begin{codebox}
\small
\begin{lstlisting}
...
========== ASSOCIATIONS ==========
Factory "1" -- "*" Machine : contains
Machine "1" -- "*" ProductType : produces
Machine "*" -- "*" Worker : operates
Worker "*" -- "*" Machine : works in
OrderLine "1" -- "1" ProductType : specifies
PurchaseOrder "1" -- "*" OrderLine : contains
Client "1" -- "*" PurchaseOrder : issues
...
\end{lstlisting}
\end{codebox}
\caption{B2 Fabricated Code and B3 Contradiction. The PlantUML code introduces \texttt{Factory--Machine} and \texttt{Machine--Worker} relations absent from the reasoning plan (B2), and changes the multiplicity and direction of \texttt{OrderLine--ProductType} (B3).}
\label{fig:ex:b2b3}
\end{figure}
 
\paragraph{B3, Contradiction.} The code conflicts with the plan on multiplicity, relation kind/direction, or type. In the same example (Figure~\ref{fig:ex:b2b3}), the reasoning specifies \texttt{OrderLine "*" -{}-{}> "1" ProductType} but the code implements \texttt{OrderLine "1" -{}- "1" ProductType}, changing both multiplicity and direction.
 
\paragraph{B4, Structural Mismatch.} The reasoning trace 
specifies one structural choice but the PlantUML code 
implements another (e.g., a planned association realised as 
an attribute/collection; aggregation vs.\ composition; 
interface planned but class emitted). This is not a simple 
Dropped Step~(B1) nor a Contradiction~(B3), but a change of 
modelling construct.
 
Figure~\ref{fig:ex:b4} illustrates this. The reasoning trace proposes a separate \texttt{StockItem} class that references \texttt{Movie} and carries inventory properties such as availability and pricing. In the PlantUML code, these structures are collapsed into attributes of \texttt{Movie}.

\begin{figure}[!ht]
\begin{reasonbox}
\small
 \ldots
\begin{lstlisting}
We can design the Catalogue to manage the StockItems.
...
Movie
- id: int
int
- title: String
- director: String
- year: int
 
 
StockItem
- id: int
- movie: Movie
- available: boolean
- buyPrice: double
- rentPrice: double
 
\end{lstlisting}
\end{reasonbox}
 
\begin{codebox}
\small
\begin{lstlisting}
 
class Movie {
- movieId: int
- title: String
- director: String
- releaseYear: int
- isAvailable: boolean
}
 
\end{lstlisting}
\end{codebox}
\caption{B4 Structural Mismatch. The reasoning trace plans a separate \texttt{StockItem} class with inventory properties, but the PlantUML code collapses these into attributes of \texttt{Movie}.}
\label{fig:ex:b4}
\end{figure}

\subsubsection{Dimension C: Reasoning Quality (Within-Chain Soundness)}
 
This dimension evaluates whether the reasoning trace itself is logically sound, regardless of whether errors propagate to the final code.

\paragraph{C1, Speculative Branch.}We tag Speculative Branch when the reasoning trace introduces hedged alternatives (``maybe/possibly/if needed'') that add entities, relations, 
or assumptions without citing evidence from the requirement description. These alternatives are typically abandoned or 
never resolved. If a speculative branch is explicitly retracted with a rationale and leaves no unsupported 
commitment, it is recorded but considered lower severity. However, if the speculation propagates into later design decisions or forces reversals downstream, it is scored as a 
significant fault.
 
Figure~\ref{fig:ex:c1} illustrates this. The reasoning trace introduces a possible \texttt{Factory} entity and a one-to-many relation with \texttt{Machine}, even though the requirement description provides no explicit evidence for either. The model repeatedly speculates about whether to include or omit the \texttt{Factory} class, changing its decision several times before discarding it. This behaviour represents a Speculative Branch because the model explores unsupported alternatives without clear justification. Although the speculation is eventually withdrawn, it reflects unstable reasoning that can lead to inconsistencies in later steps.

\begin{figure}[!ht]
\begin{reasonbox}
\small
 
\textcolor{orange}{
1. Factory: 
- We don't have specific attributes for the factory, but it contains machines.}
 
\ldots 
 
\textcolor{violet}{
1. Factory and Machine: 
- A factory has several machines -- one-to-many from Factory to Machine? But note: the problem does not say multiple factories? 
However, the problem states "a factory", so we are modeling one factory? Or should we consider multiple? 
Since it's a database for a factory, we can assume one factory. So we don't need a factory entity? 
Alternatively, we can assume that the system may be extended to multiple factories? 
The problem says "a factory", so we can model without a Factory entity and just have the machines as independent. 
But note: the requirement says "a factory has several machines", so if we are only dealing with one factory, we can omit the Factory class. }
 
\textcolor{magenta}{
However, for extensibility, we can have a Factory class and then have a one-to-many relationship from Factory to Machine.}
 
\textcolor{blue}{
Since the problem does not require storing multiple factories and doesn't specify attributes for the factory, we can omit it for simplicity.
}
\end{reasonbox}
\caption{C1 Speculative Branch. The reasoning trace reverses its decision on whether to include a \texttt{Factory} class multiple times, without justification from the requirement description. Each coloured fragment marks a reversal.}
\label{fig:ex:c1}
\end{figure}
 
The model oscillates multiple times about whether to include a \texttt{Factory} class without resolving the question with textual evidence.

\paragraph{C2, Goal Drift.} We record Goal Drift when the 
reasoning trace optimises for aesthetics, simplicity, or 
notational preferences (e.g., preferring symmetry, minimising 
class count, choosing composition ``for neatness'') rather 
than faithfully implementing the requirement description. The 
unit of analysis is a design decision whose stated rationale 
is misaligned with the task objective.
 
Figure~\ref{fig:ex:c2} illustrates this. The reasoning trace diverts from the task objective by focusing on the visual appearance of the diagram rather than the correctness of its structure. The model introduces aesthetic adjustments, such as the use of \texttt{skinparam} settings and crow's foot notation, which are unrelated to the functional or semantic requirements described in the requirement description.

\begin{figure}[!ht]
\begin{reasonbox}
\small
\ldots But note: the Appointment\_Condition and 
Breed\_Condition are many-to-many association tables. We can 
represent them as classes or as associations with a cross. 
In PlantUML, we can use the following for many-to-many:
 
\texttt{Appointment \}o-{}-o\{ Condition : "detects"}
 
\texttt{Breed \}o-{}-o\{ Condition : "common for"}
 
\ldots
 
Use skinparam to make it look better
 
\texttt{skinparam linetype ortho}
 
\ldots But note: the above uses crow's foot notation. Let me 
explain:
 
- Owner to Animal: one owner to zero or more animals (the 
'o' means optional, i.e., 0..*)
 
- Similarly, Breed to Animal: one breed to zero or more 
animals.
 
\ldots
 
- Appointment to Condition: many-to-many (we use 
\texttt{\}o-{}-||} for many-to-many? Actually, in crow's 
foot, we use:
 
\quad - \texttt{||} for one
 
\quad - \texttt{\}o} for many (zero or more)
\end{reasonbox}
\caption{C2 Goal Drift. The reasoning trace shifts focus to visual presentation (\texttt{skinparam}, crow's-foot notation) rather than faithful implementation of the requirement.}
\label{fig:ex:c2}
\end{figure}

\paragraph{C3, Tool Misuse.} We label Tool Misuse when the reasoning trace proposes PlantUML or UML constructs with incorrect semantics or syntax, such as composition without lifecycle dependency, using inheritance to model a role, misusing realisation/interfaces, or invalid multiplicities. 
Faults are counted even if the final PlantUML code omits them, since proposing an invalid construct indicates a 
misunderstanding that may reappear as contradictions or structural substitutions in the code.
 
Figure~\ref{fig:ex:c3} illustrates this. The reasoning trace intends to represent a composition relationship between \texttt{Order} and \texttt{OrderLine} but uses the incorrect PlantUML notation \texttt{-{}-o}. In PlantUML, the symbol \texttt{o-{}-} denotes aggregation, not composition.

\begin{figure}[!ht]
\begin{reasonbox}
\small
\texttt{' Order and OrderLine: \textcolor{violet}{composition}, because the lines are part of the order}

\texttt{Order "1" -{}-o "1..*" OrderLine : consists of >}
\end{reasonbox}
\caption{C3 Tool Misuse. The reasoning trace intends composition between \texttt{Order} and \texttt{OrderLine} but uses \texttt{-{}-o}, the PlantUML symbol for aggregation.}
\label{fig:ex:c3}
\end{figure}
 
The reasoning intends composition but uses the PlantUML notation \texttt{-{}-o}, which denotes aggregation. This indicates a misunderstanding of the notation that may propagate to the final output.

\subsection{RQ2: Hallucination Distribution}\label{sec:RQ2:HallDist}
 
\begin{figure}[t]
  \centering
  \includegraphics[width=\columnwidth]{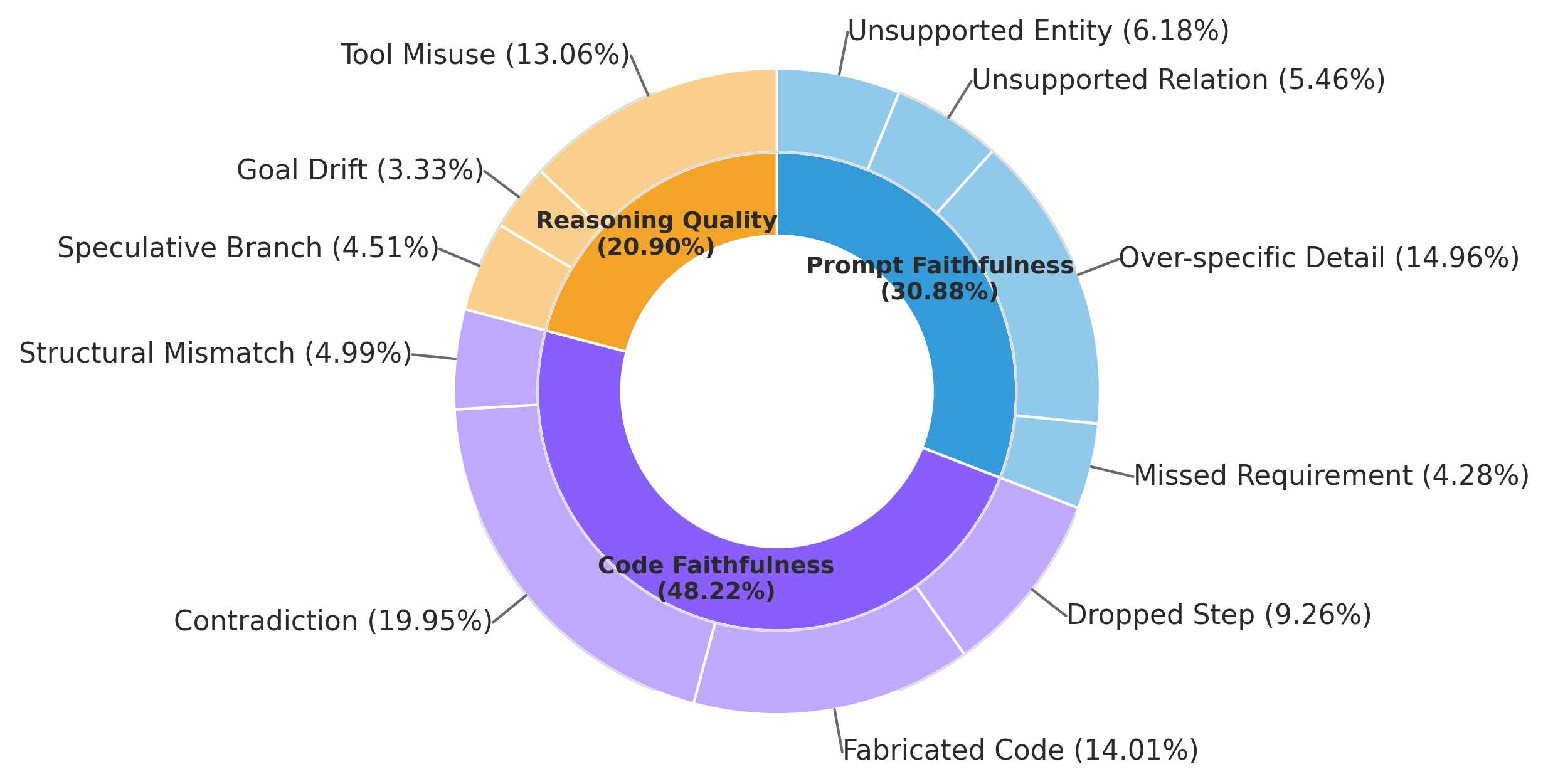}
  \caption{Proportional distribution of hallucination types in Dataset~1 (60 manually annotated runs).}
  \label{fig:dist1}
\end{figure}

Figure~\ref{fig:dist1} shows the proportional distribution of hallucination types across the 60 runs of Dataset~1 (manual annotation). Code Faithfulness was the most frequent dimension (48.2\%), followed by Prompt Faithfulness (30.9\%) and Reasoning Quality (20.9\%). Within Code Faithfulness, the leading sub-types were Contradiction (20.0\%) and Fabricated Code (14.0\%). Over-specific Detail (15.0\%) was the leading sub-type within Prompt Faithfulness, accounting for nearly half of all errors in that dimension. Tool Misuse (13.1\%) clearly dominated Reasoning Quality, where it constituted 62.5\% of within-dimension errors; Speculative Branch (4.5\%) and Goal Drift (3.3\%) appeared far less frequently.
 
\begin{figure}[t]
  \centering
  \includegraphics[width=\columnwidth]{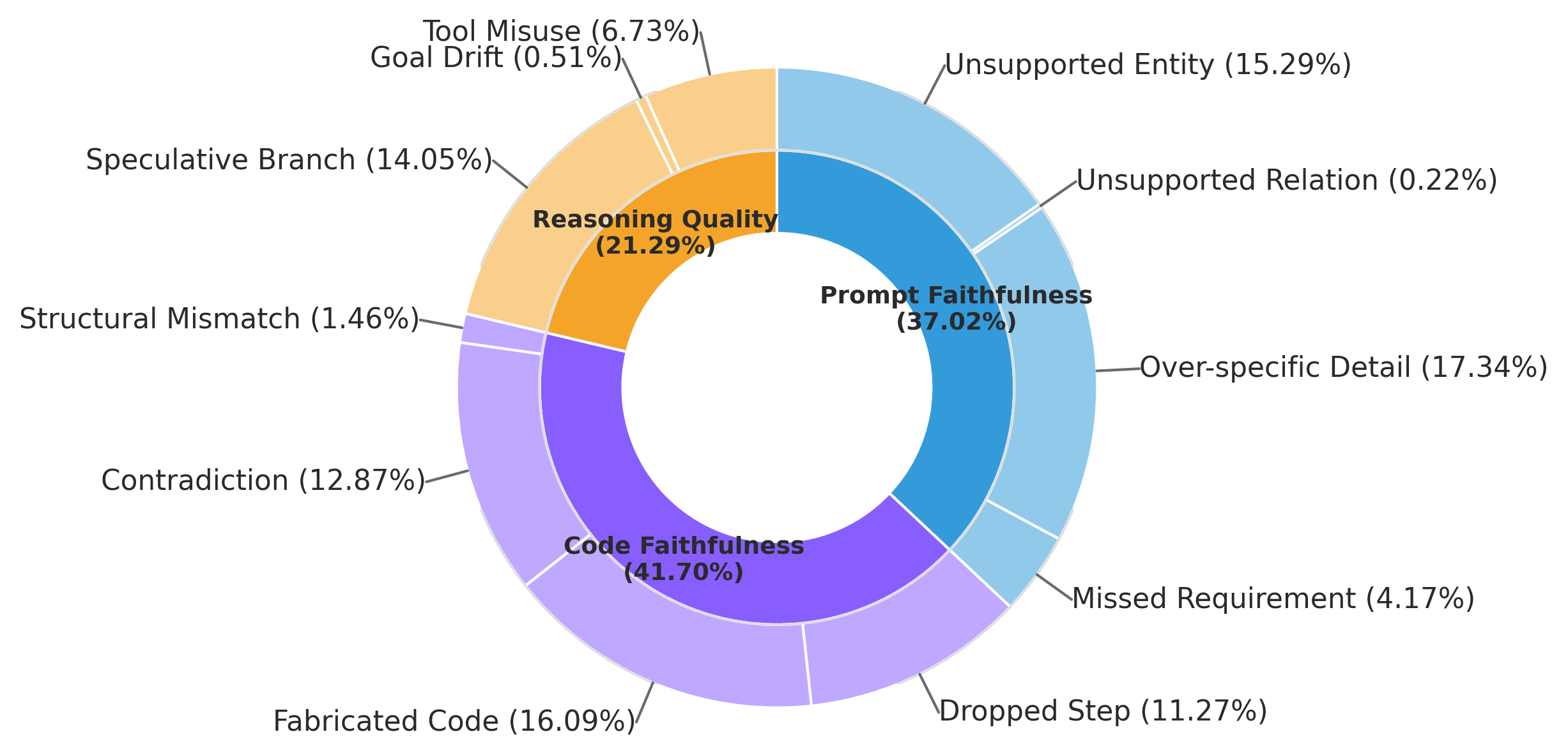}
  \caption{Proportional distribution of hallucination types in Dataset~2 (120 descriptions, automated dual-LLM annotation).}
  \label{fig:dist2}
\end{figure}

Applying the dual-LLM evaluation pipeline (Section~\ref{sec:design_study2}) to the 120 descriptions in Dataset~2 yielded the distribution shown in Figure~\ref{fig:dist2}. The dimension-level ordering was consistent with Dataset~1: Code Faithfulness remained the most frequent dimension (41.7\% vs.~48.2\%), followed by Prompt Faithfulness (37.0\% vs.~30.9\%) and Reasoning Quality (21.3\% vs.~20.9\%). Sub-type rankings were also partially consistent: Fabricated Code and Contradiction were the most frequent Code Faithfulness sub-types in both datasets, while Over-specific Detail led Prompt Faithfulness. For Reasoning Quality, however, the leading sub-type differed across annotation settings, with Tool Misuse dominating Dataset~1 and Speculative Branch dominating Dataset~2. Automated annotation also yielded higher absolute error counts than manual coding.

To assess the consistency of the dual-LLM annotation pipeline
with human judgement, one author independently annotated a
stratified random sample of 30 reasoning traces (25\% of
Dataset~2). We compared hallucination instances identified
by the pipeline and the human annotator using pairwise F1,
considering an instance matched when both identified the
same hallucination at the same location in the trace. Pairwise F1 was 77.93\% for Prompt Faithfulness ($n=130$),
87.60\% for Code Faithfulness ($n=145$), and 81.60\% for
Reasoning Quality ($n=74$), with an overall micro-averaged
F1 of 82.89\% ($N=349$). Here, $n$ denotes the number of distinct hallucination instances
identified by at least one of the two annotation methods within
each dimension, while $N$ denotes the total across all three dimensions.

\begin{figure}[t]
  \centering
  \includegraphics[width=\columnwidth]{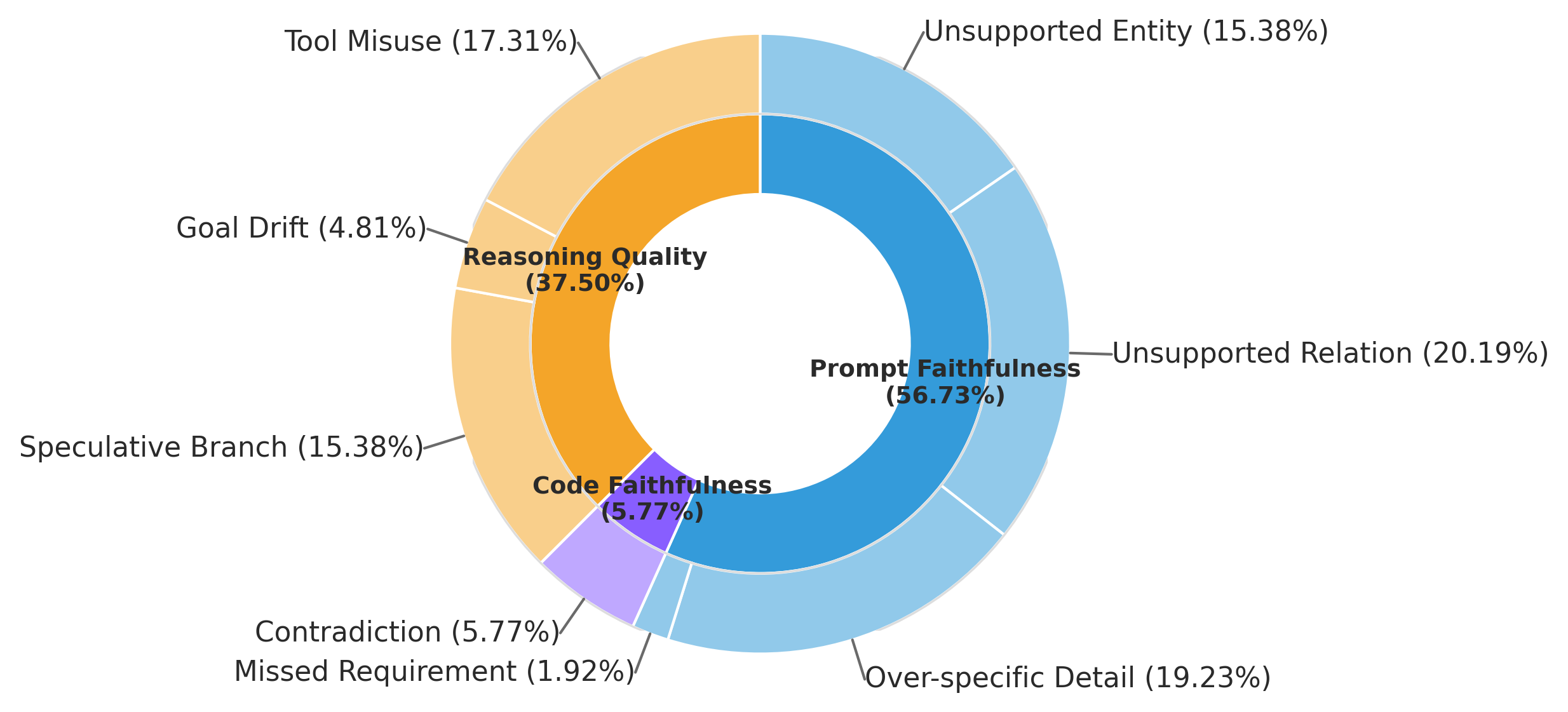}
\caption{Proportional distribution of hallucination types for DeepSeek-V4-Pro (20 manually annotated traces on the Dataset~1 descriptions).}
  \label{fig:dist_v4}
\end{figure}

To assess whether the proposed taxonomy generalises beyond the model used for its initial construction, we repeated the procedure on all 20 descriptions from Dataset~1 using the more recent DeepSeek-V4-Pro model at temperature~1. The resulting 20 reasoning traces were annotated manually following the procedure in Section~\ref{sec:design_study1}. The taxonomy remained applicable to the new traces, but the distribution of hallucination dimensions differed substantially from that observed for DeepSeek-R1 on the same descriptions (Table~\ref{tab:dist_comparison} and Figure~\ref{fig:dist_v4}). Prompt Faithfulness became the most frequent dimension (56.7\%), followed by Reasoning Quality (37.5\%), while Code Faithfulness decreased sharply to 5.8\%. All six Code Faithfulness instances were classified as Contradiction (B3), corresponding specifically to multiplicity reversals between the reasoning trace and the generated code. Within Prompt Faithfulness, Unsupported Relation (A2; 20.2\% of all hallucinations) and Over-specific Detail (A3; 19.2\%) were the two most frequent sub-types. Over-specific Detail thus remained among the leading Prompt Faithfulness sub-types, as in Datasets~1 and~2, whereas Unsupported Relation, which accounted for only 5.5\% of hallucinations in Dataset~1, became considerably more prominent.

\begin{table}[t]
\caption{Dimension-level hallucination distribution across datasets and models. 
DeepSeek-V4-Pro results are based on 20 traces and should be interpreted 
with caution.}
\label{tab:dist_comparison}
\centering
\small
\begin{tabular}{lccc}
\hline\noalign{\smallskip}
Dimension & Dataset~1 & Dataset~2 & DeepSeek-V4-Pro \\
          & (R1, $n=60$) & ($n=120$) & ($n=20$) \\
\noalign{\smallskip}\hline\noalign{\smallskip}
Prompt Faithfulness  & 30.9\% & 37.0\% & 56.7\% \\
Code Faithfulness    & 48.2\% & 41.7\% &  5.8\% \\
Reasoning Quality    & 20.9\% & 21.3\% & 37.5\% \\
\noalign{\smallskip}\hline
\end{tabular}
\end{table}

\subsection{RQ3: Reasoning Length and Requirement Complexity}

\begin{figure}[t]
  \centering
  \includegraphics[width=0.85\columnwidth]{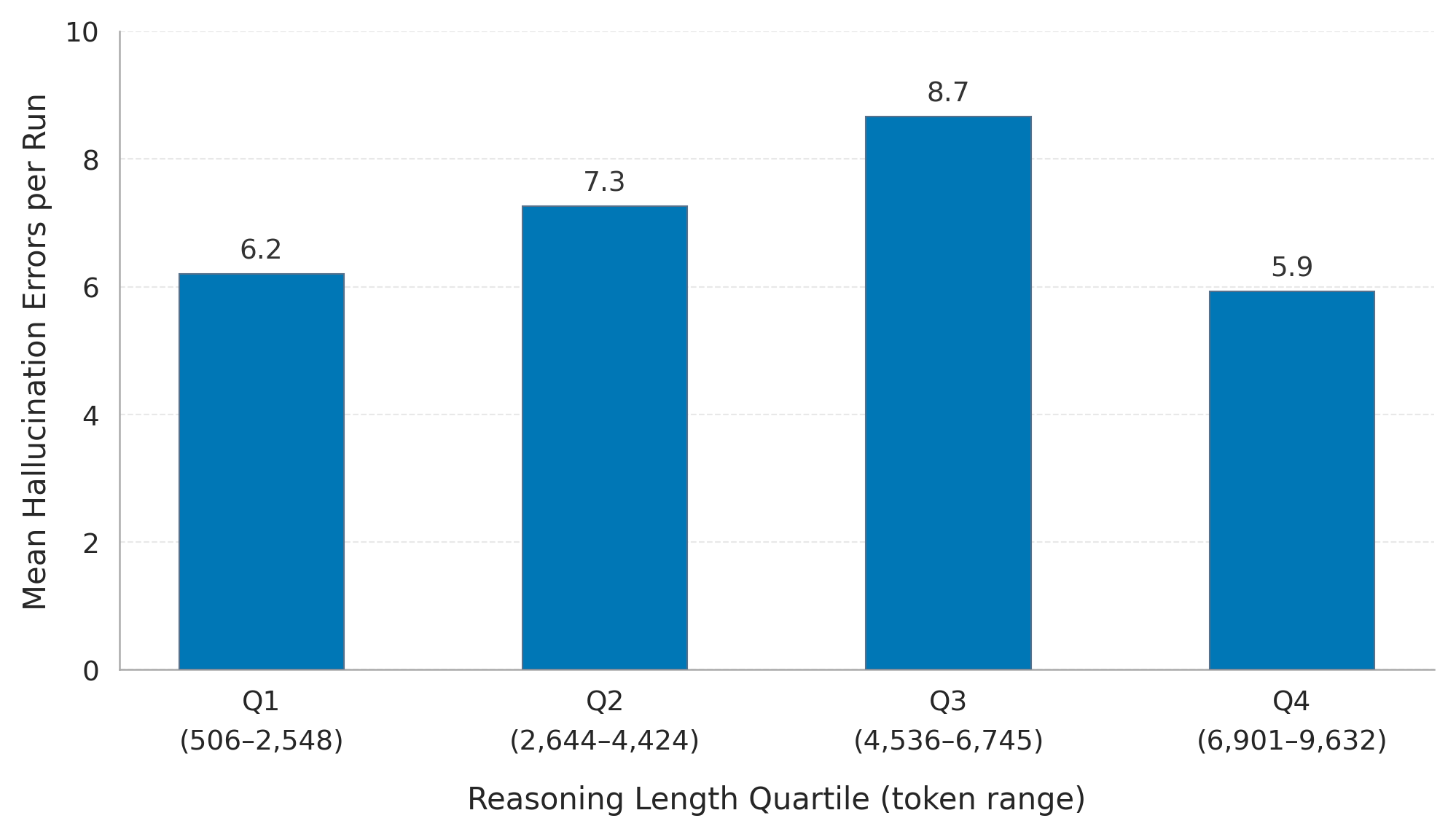}
  \caption{Mean total errors per run by reasoning-length quartile (Dataset~1, $N = 60$).}
  
  \label{fig:rq3_length}
\end{figure}

Using Dataset~1, we examined the association between
reasoning-trace length (in tokens) and the number of hallucination
errors per run. Figure~\ref{fig:rq3_length} presents the mean
total errors across four quartiles of reasoning length
($n=15$ runs per quartile). Descriptively, the binned means
suggest an inverted-U pattern, with fewer errors in the
shortest (Q1: 6.20) and longest (Q4: 5.93) quartiles and
higher error counts in the middle quartiles (Q2: 7.27;
Q3: 8.67). However, Spearman correlation analyses found
no statistically significant monotonic association between
reasoning length and overall hallucination frequency
($\rho=-0.011$, $p=0.934$), or any individual hallucination
dimension (all $p>0.14$). Thus, the inverted-U pattern
should be interpreted as a descriptive tendency rather
than evidence of a statistically established relationship.

We examined the association
between requirement complexity, computed using
Equation~\eqref{eq:complexity}, and hallucination frequency.
Requirement complexity showed a negative correlation
with overall hallucination frequency (Spearman
$\rho=-0.325$, $p=0.011$), indicating that runs with
higher requirement-complexity scores tended to exhibit
fewer annotated hallucinations. This association was
most pronounced for Prompt Faithfulness
($\rho=-0.375$, $p=0.003$), while correlations with
Reasoning Quality ($\rho=-0.209$, $p=0.110$) and
Code Faithfulness ($\rho=-0.173$, $p=0.187$)
were not statistically significant. These findings
suggest a modest association between requirement
complexity and hallucination frequency, particularly
for deviations from the original requirements.
However, the correlational analysis does not
establish a causal relationship.

\section{Taxonomy-Guided Mitigation Experiment}\label{sec:TaxMitigation}
 
To answer RQ4, we evaluated whether embedding the taxonomy as generation-time constraints and post-hoc verification rules can improve diagram quality.
 
Three conditions were designed to isolate the effects of embedding different taxonomy dimensions into the generation pipeline.
 
\emph{Condition~1, Baseline:} A plain system prompt with no taxonomy awareness and no post-hoc verification. The model receives a minimal instruction (``You are a helpful assistant that creates UML class diagrams'') and generates PlantUML code from the requirement description. This serves as the control.
 
\emph{Condition~2, Taxonomy-Aware Generation:} The system prompt encodes two taxonomy dimensions as explicit generation constraints: Prompt Faithfulness (A1--A4) and Reasoning Quality (C1--C3). The model is instructed to identify every entity, action, and constraint in the requirement; justify each class with an explicit quotation from the text; assign multiplicities only when textually supported; and re-read the requirement to verify completeness before finalising. Key constraints prohibit introducing unsupported entities or relations (A1, A2), inventing datatypes or multiplicities (A3), using composition without lifecycle dependency (C3), and speculative branching (C1).
 
\emph{Condition~3, Taxonomy-Aware Generation + Code Faithfulness Review:} This condition reuses the generation output from Condition~2 and adds a post-hoc verification phase that operationalises the remaining taxonomy dimension: Code Faithfulness (B1--B4). A second prompt provides the model with the full reasoning trace and the generated PlantUML code, instructing it to detect Dropped Steps~(B1), Fabricated Code~(B2), Contradictions~(B3), and Structural Mismatches~(B4). Both the generation and verification phases use thinking mode.
 
Three models were evaluated to assess whether taxonomy-guided prompting generalises across model sizes and training variants: Qwen3-32B (a large-scale instruction-tuned model), Qwen3-8B (a smaller variant), and DeepSeek-R1-0528-Qwen3-8B (hereafter, DeepSeek-R1-Qwen3-8B), a distilled model specialised for reasoning tasks. All models were run with the temperature settings recommended by Hugging Face and with thinking mode enabled.
 
Generated diagrams were evaluated against the gold-standard PlantUML solutions from Dataset~3 along the two semantic-quality dimensions of \emph{completeness} and \emph{correctness} introduced by ~\cite{lindland1994understanding} in their framework for conceptual-modelling quality. We adapt these dimensions to the UML class-diagram setting by decomposing each into three structural categories, classes, attributes, and relationships and scoring every category on a 0--100\% scale. This yields six per-diagram metrics: three completeness (recall) measures (i.e., Class Completeness, Attribute Completeness, and Relationship Completeness) and three correctness (precision) measures (i.e., Class Correctness, Attribute Correctness, and Relationship Correctness). The gold solution served as the primary reference; the natural-language description was used as a secondary reference to validate whether deviations were semantically justified. Evaluation scores were independently reviewed and verified by two authors.

To provide a balanced single-metric view of diagram quality, we combine completeness and correctness into an $F_1$ score per dimension:
\begin{equation}
\label{eq:f1}
F_1 = \frac{2 \times \text{Correctness} \times \text{Completeness}}{\text{Correctness} + \text{Completeness}}
\end{equation}
We compute $F_1$ for each of the three dimensions (Class, Attribute, Relationship) and report their average as the Overall~$F_1$. This metric penalises both hallucinated elements (low precision) and missed requirements (low recall).

 %%%%%% Second Section

Table~\ref{tab:rq4_f1} presents the mean $F_1$ scores across all 13 Golden UML ModelSet cases for each model and condition. Table~\ref{tab:rq4_deltas} reports the change relative to the Baseline with statistical significance.
 
\begin{table}[t]
\caption{Mean $F_1$ scores (\%) per model and condition across 13 cases. Bold indicates the best score per model and metric.}
\label{tab:rq4_f1}
\centering
\footnotesize
\begin{tabularx}{\textwidth}{lXcccc}
\hline\noalign{\smallskip}
Model & Condition & Cls & Att & Rel & Overall \\
\noalign{\smallskip}\hline\noalign{\smallskip}
\multirow{3}{*}{\shortstack[l]{DeepSeek-R1-Qwen3-8B}}
  & Baseline            & 84.1 & 84.8 & 67.5 & 78.8 \\
  & T-Aware Gen.        & 88.2 & \textbf{89.6} & \textbf{74.3} & \textbf{84.0} \\
  & T-Aware + CF Rev.   & \textbf{88.8} & 88.7 & 71.4 & 83.0 \\
\noalign{\smallskip}\hline\noalign{\smallskip}
\multirow{3}{*}{Qwen3-32B}
  & Baseline            & 89.7 & 85.0 & 82.0 & 85.5 \\
  & T-Aware Gen.        & \textbf{90.9} & \textbf{92.5} & \textbf{86.2} & \textbf{89.9} \\
  & T-Aware + CF Rev.   & 90.2 & 90.4 & 85.1 & 88.6 \\
\noalign{\smallskip}\hline\noalign{\smallskip}
\multirow{3}{*}{Qwen3-8B}
  & Baseline            & 86.5 & 93.9 & 70.4 & 83.6 \\
  & T-Aware Gen.        & \textbf{90.6} & \textbf{94.0} & 73.9 & 86.1 \\
  & T-Aware + CF Rev.   & \textbf{90.6} & 93.3 & \textbf{76.9} & \textbf{86.9} \\
\noalign{\smallskip}\hline
\end{tabularx}
\end{table}
 
\begin{table}[t]
\caption{Change in Overall $F_1$ (pp) relative to Baseline. Significant results (Wilcoxon, $p < 0.05$) marked with~*.}
\label{tab:rq4_deltas}
\centering
\footnotesize
\begin{tabularx}{\textwidth}{lXccc}
\hline\noalign{\smallskip}
Model & Condition & $\Delta F_1$ & $W$ & $p$ \\
\noalign{\smallskip}\hline\noalign{\smallskip}
\multirow{2}{*}{\shortstack[l]{DeepSeek-R1-Qwen3-8B}}
  & T-Aware Gen.      & +5.2 & 8.0  & 0.006* \\
  & T-Aware + CF Rev. & +4.1 & 8.0  & 0.006* \\
\noalign{\smallskip}\hline\noalign{\smallskip}
\multirow{2}{*}{Qwen3-32B}
  & T-Aware Gen.      & +4.3 & 8.0  & 0.006* \\
  & T-Aware + CF Rev. & +3.0 & 15.0 & 0.064 \\
\noalign{\smallskip}\hline\noalign{\smallskip}
\multirow{2}{*}{Qwen3-8B}
  & T-Aware Gen.      & +2.6 & 16.0 & 0.040* \\
  & T-Aware + CF Rev. & +3.4 & 15.0 & 0.033* \\
\noalign{\smallskip}\hline\noalign{\smallskip}
\multicolumn{2}{l}{\emph{Pooled}} & & & \\
  & T-Aware Gen.      & +4.0 & 90.0  & $<$0.001* \\
  & T-Aware + CF Rev. & +3.5 & 116.0 & $<$0.001* \\
\noalign{\smallskip}\hline
\end{tabularx}
\end{table}
 
\paragraph{Taxonomy-aware generation improves diagram quality.}
Taxonomy-Aware Generation, which incorporates the Prompt Faithfulness and Reasoning Quality dimensions into the generation prompt, improved Overall~$F_1$ for all three models relative to the Baseline. The gains were +5.2~pp for DeepSeek-R1-Qwen3-8B ($W=8$, $p=0.006$), +4.3~pp for Qwen3-32B ($W=8$, $p=0.006$), and +2.6~pp for Qwen3-8B ($W=16$, $p=0.040$). The pooled improvement was +4.0~pp ($W=90$, $p<0.001$). Improvements were observed across all three structural metrics, although the largest gain differed by model: Relationship~$F_1$ improved most for DeepSeek-R1-Qwen3-8B (+6.8~pp), Attribute~$F_1$ for Qwen3-32B (+7.5~pp), and Class~$F_1$ for Qwen3-8B (+4.1~pp). These results show that Taxonomy-Aware Generation improved the structural quality of generated diagrams across the three evaluated models.

Adding the Code Faithfulness review after Taxonomy-Aware Generation produced inconsistent changes in Overall~$F_1$ across the three models. Overall~$F_1$ decreased by 1.0~pp for DeepSeek-R1-Qwen3-8B and by 1.3~pp for Qwen3-32B, while increasing by 0.8~pp for Qwen3-8B. Nevertheless, both Taxonomy-Aware Generation and Taxonomy-Aware Generation with Code Faithfulness review showed significant improvements over the Baseline in the pooled analysis (+4.0~pp and +3.5~pp, respectively; both $p<0.001$). 
After Bonferroni correction for the six per-model
comparisons against the Baseline, only three
remained statistically significant.

Relationship~$F_1$ was consistently the lowest of the
three evaluated metrics across all models and conditions,
ranging from 67.5\% for DeepSeek-R1-Qwen3-8B under the
Baseline to 86.2\% for Qwen3-32B under Taxonomy-Aware
Generation.

\section{Discussion}
\label{sec:discussion}

The four research questions form a progression from diagnosis
(RQ1), through validation (RQ2) and predictor analysis (RQ3),
to intervention (RQ4). Taken together, the findings suggest
three main conclusions.

First, Code Faithfulness was the most frequently observed
failure dimension for DeepSeek-R1, highlighting discrepancies
between reasoning and generated diagram code. However, the
distribution of hallucination dimensions differed for
DeepSeek-V4-Pro, suggesting that the relative prevalence of
failure types may depend on the model. Second, the relationships between hallucination frequency
and the examined characteristics were mixed. Reasoning-trace
length showed no significant monotonic association with
hallucination frequency, whereas requirement complexity
was negatively associated with overall hallucination
frequency and Prompt Faithfulness errors in Dataset~1. Third, incorporating taxonomy-derived guidance into
generation prompts improved mean overall F1 across all
three evaluated models, although the statistical evidence
varied across model-condition comparisons. Notably,
these improvements were obtained through guidance targeting
Prompt Faithfulness and Reasoning Quality, despite Code
Faithfulness being the most frequent failure dimension
for DeepSeek-R1. This suggests that the most frequently
observed failure dimension is not necessarily the only
useful target for intervention.

\paragraph{Localising hallucinations across the generation process.}
Code Faithfulness was the most frequent hallucination dimension
in both Dataset~1 (48.2\%) and Dataset~2 (41.7\%), with
Contradiction and Fabricated Code among its leading sub-types.
The same dimension-level ordering (Code Faithfulness, followed
by Prompt Faithfulness and Reasoning Quality) was observed
in both the manually and automatically annotated datasets,
indicating a consistent descriptive pattern across the two
datasets and annotation approaches.

For DeepSeek-R1, this highlights the importance of discrepancies
between the reasoning trace and the generated PlantUML code:
a model may formulate a plausible structural plan yet fail
to represent it faithfully in the final diagram code.
However, because the traces in both datasets were produced
by the same model, this ordering should not be assumed to
generalise to other models. Indeed, the cross-model analysis
(Section~\ref{sec:RQ2:HallDist}) indicates a different
distribution of hallucination dimensions.

\paragraph{Failure distributions may be model-dependent.}
The cross-model analysis qualifies the pattern above. For DeepSeek-V4-Pro, Code Faithfulness accounted for only 5.8\% of observed hallucinations, all involving multiplicity reversals between the reasoning trace and the generated code, while Prompt Faithfulness (56.7\%) and Reasoning Quality (37.5\%) became the dominant dimensions. The taxonomy remained applicable to the new traces, suggesting that it captures failure modes that recur across models, although their relative prevalence may differ substantially. In particular, the reasoning-to-code transition, which was the dominant failure site for DeepSeek-R1, showed substantially fewer observed failures for DeepSeek-V4-Pro, although hallucinations were not eliminated overall. The persistence of Over-specific Detail as the leading Prompt Faithfulness sub-type across the analysed datasets and models suggests that some individual failure patterns may be more stable than their aggregate frequencies. Because the cross-model analysis is based on only 20 traces at a single temperature, these findings provide preliminary evidence of transferability and should be confirmed through a more systematic 
cross-model evaluation.

\paragraph{Reasoning length and requirement complexity show
different associations with hallucination frequency.}
In Dataset~1, reasoning-trace length showed no significant
monotonic association with hallucination frequency,
suggesting that longer reasoning does not necessarily
correspond to more or fewer observed errors. Although
the quartile-based analysis suggested an inverted-U
pattern, this remained a descriptive observation rather
than a statistically established relationship.

Requirement complexity, in contrast, was negatively
associated with overall hallucination frequency,
particularly with Prompt Faithfulness errors. One
possible explanation is that requirements receiving
higher complexity scores may also provide more explicit
information that constrains unsupported inferences.
However, this interpretation remains speculative,
as the complexity measure does not independently
capture requirement clarity or specificity.

Overall, these findings suggest that simple measures
such as reasoning length and requirement complexity
are insufficient on their own to characterise
hallucination frequency. Given that the analysis
was restricted to Dataset~1, further investigation
is needed to determine whether these associations
hold across other datasets and models.

\paragraph{From failure analysis to intervention.}
Taxonomy-Aware Generation improved Overall~$F_1$ across all three evaluated models, showing that the taxonomy can be operationalised beyond descriptive failure analysis and used to inform a generation-time intervention. This provides an important connection between the diagnostic and mitigation stages of the study: failure patterns identified through the analysis of reasoning traces were translated into generation constraints and evaluated through the quality of the resulting diagrams. 

The post-hoc Code Faithfulness review produced a different pattern. Adding the review to Taxonomy-Aware Generation decreased mean Overall~$F_1$ for two models and increased it for one, providing no consistent additional benefit. This is noteworthy because Code Faithfulness was the most frequent hallucination dimension in the initial DeepSeek-R1 analyses. One possible explanation is that correcting consistency between the reasoning trace and generated code does not necessarily improve the final diagram when the underlying reasoning is itself flawed. Some Code Faithfulness errors may also reflect earlier failures in the generation process, although this mechanism was not directly tested. The review added in Condition~3 differs from the constraints in Condition~2 in both its timing and the taxonomy dimension it targets. Therefore, the observed difference cannot be attributed to intervention timing alone and does not establish that generation-time constraints are generally more effective than post-hoc verification. Alternative forms of post-hoc verification may produce different results.

\paragraph{Relationship modelling remains a persistent challenge.}
\paragraph{Relationship modelling remains a persistent challenge.}
Relationship~$F_1$ was the lowest-scoring metric for every model and condition (Table~\ref{tab:rq4_f1}). Although taxonomy-aware generation improved Relationship~$F_1$ for all three models, the magnitude of improvement varied, and relationships did not consistently show the largest gains among the evaluated diagram elements. Relationship modelling therefore remains an important target for further improvement.

Two characteristics of UML relationships may help explain
this difficulty. First, natural-l   anguage requirements
may leave relationship properties implicit or
underspecified. For example, a statement such as
\emph{``students enrol in courses''} identifies related
entities but does not necessarily specify multiplicity,
navigability, or whether enrolment should be modelled
as a direct association or an association class.
Second, identifying appropriate relationships may
require integrating information distributed across
multiple requirements rather than relying on local
textual matches. Although these explanations were
not directly tested, they suggest potential directions
for future interventions, including few-shot examples
of common relationship patterns, explicit relationship
inventories, and dedicated multiplicity verification.

\paragraph{Prompt echoing as a potential diagnostic cue.} One qualitative observation from the manual coding of Dataset~1 is worth noting alongside the quantitative findings: a recurring \emph{prompt-echoing} pattern in which the model restates a span of the requirement verbatim within its reasoning trace before committing to a structural decision. Echoing was typically signalled by phrases such as ``the problem says\ldots'', ``the description states that\ldots'', or ``it says\ldots''. In our coded traces it appeared to function as a boundary marker between grounded reasoning and speculative commitment: the model frequently used the echo as an apparent justification for a structural decision that the prompt did not in fact support, treating the recitation itself as evidence. Figure~\ref{fig:echo} shows a representative case from the project-modelling exercise, in which the model echoes ``work products are created, used, updated, and elaborated throughout the project'' and then commits to composition between Project and its Requirements/System. The resulting commitment was coded as Unsupported Relation (A2): the prompt's verb ``produces'' licenses a general association but not the lifecycle dependency that composition implies. Operationally, echo-then-commit transitions are mechanically simple to detect---a sliding-window match between the reasoning trace and the prompt---and could be incorporated into existing LLM observability pipelines to flag candidate hallucinations for review.
 
\begin{figure}[!ht]
\begin{echobox}
\small
However, the problem says ``work products are created, used, updated, and elaborated throughout the project''. So both requirements and system are work products of the project. So composition is appropriate.
 
But the project has multiple requirements and one system. We'll use composition for both:
 
\texttt{Project "1" *-{}- "*" Requirement}
 
\texttt{Project "1" *-{}- "1" System}
\end{echobox}
\caption{The prompt-echoing anti-pattern. The reasoning trace recites a span of the requirement verbatim and then commits to composition between Project and its Requirements/System. The commitment was coded as A2 Unsupported Relation: the prompt's verb ``produces'' supports a general association but not the lifecycle dependency composition implies.}
\label{fig:echo}
\end{figure}
 
\paragraph{Implications for practice.}
Two practical implications follow from our findings for practitioners using reasoning-capable LLMs to generate UML class diagrams. First, incorporating modelling-specific constraints into the generation prompt can be useful: across all three evaluated models, the Taxonomy-Aware prompt improved diagram quality relative to the minimal Baseline prompt. Second, relationship modelling warrants particular attention during generation and review. Relationship~$F_1$ was the lowest structural metric across all models and conditions, including after Taxonomy-Aware Generation. Practitioners may therefore benefit from explicitly checking generated relationships and multiplicities against the source requirements rather than relying only on the overall plausibility of the diagram. More broadly, our findings suggest that improving LLM-based UML generation may benefit from modelling-specific mechanisms for grounding and validating structural decisions alongside general reasoning capabilities.

\section{Threats to Validity}
\label{sec:threats}
 We discuss the main threats to the validity of our study below.

\textbf{Internal Validity.}
The taxonomy was developed through manual coding and
subsequent independent annotation, with disagreements
resolved through discussion and consensus. Researcher
interpretation may nevertheless have influenced category
boundaries and the classification of borderline cases.
Although pre-consensus consistency was assessed using
pairwise F1, the lower agreement observed for Reasoning
Quality indicates greater annotation uncertainty within
this dimension. A further limitation concerns the interpretation of
reasoning traces. Our analysis treats the emitted reasoning
text as observable evidence of the model's reasoning
process. However, these traces may not faithfully or
completely represent the model's underlying computation.
Consequently, the identified hallucination patterns
should be understood as discrepancies observable in
the generated traces rather than definitive evidence
of the model's internal decision-making mechanisms.

\textbf{External Validity.}
The datasets used in this study primarily comprise
educational and domain-modelling tasks and may not
fully represent the linguistic or structural diversity
of industrial requirements. The Golden UML
ModelSet~\cite{verbruggen2025golden}, used in the
mitigation experiment (RQ4), covers 13 modelling
domains but remains rooted in educational contexts.

The taxonomy was developed from DeepSeek-R1 reasoning
traces. Although its applicability was also examined
using traces from DeepSeek-V4-Pro, this cross-model
analysis involved only 20 traces from another model
in the same family. Furthermore, the mitigation
experiment evaluated three Qwen-based models,
including a DeepSeek-R1-distilled variant.
Consequently, the generalisability of the taxonomy
and mitigation results to other model architectures
and non-reasoning LLMs remains uncertain.

Finally, the mitigation experiment included only
13 cases per condition. Although these cases cover
the full set of domains represented in the selected
benchmark, the limited sample size restricts the
extent to which the observed improvements can be
generalised to broader modelling tasks.

\textbf{Construct Validity.}
The taxonomy focuses on hallucinations in requirements
interpretation, reasoning, and generated code, and does
not cover all aspects of diagram quality, such as layout
and readability. Pre-consensus pairwise~$F_1$ was lowest
for Reasoning Quality, indicating greater annotation
uncertainty in this dimension. The dual-LLM annotation pipeline may introduce systematic model biases. To mitigate dependence on a single provider, the annotator and judge were selected from different providers. Nevertheless, this separation does not eliminate the possibility of shared biases or
classification errors.

\textbf{Conclusion Validity.}
The mitigation experiment (RQ4) involved only 13 paired
cases per model, limiting the statistical power of
the Wilcoxon signed-rank tests. Although several
improvements were statistically significant, some
did not remain significant after Bonferroni correction
for multiple comparisons. Non-significant results
should therefore not be interpreted as evidence
of no effect. The correlation analyses (RQ3) were restricted to
60 reasoning traces from Dataset~1, limiting
sensitivity to small associations. Furthermore,
the observed correlations should be interpreted
as associations rather than causal relationships.
These findings require confirmation using larger
and more diverse samples.

\section{Conclusion}
\label{sec:conclusion}

This paper investigated hallucinations in UML class-diagram generation by reasoning-capable LLMs, moving beyond the observation of structural errors to characterise how and where they arise during generation. From manually coded reasoning traces and large-scale automated validation, we developed a taxonomy comprising three dimensions: Prompt Faithfulness, Code Faithfulness, and Reasoning Quality. These dimensions correspond to distinct stages of the requirements-to-reasoning-to-code process.

Our results show that Code Faithfulness is the most frequent failure dimension in both DeepSeek-R1 datasets, but that the distribution of hallucinations varies across models. Reasoning-trace length showed no significant association with hallucination frequency, while requirement complexity was negatively associated with overall hallucination frequency and Prompt Faithfulness errors. Most importantly, the taxonomy is actionable: incorporating its Prompt Faithfulness and Reasoning Quality dimensions into the generation prompt increased pooled mean Overall~$F_1$ by 4.0~pp across the evaluated models, while an additional post-hoc Code Faithfulness review provided no consistent benefit. This suggests that understanding the origins of hallucinations can inform interventions aimed at preventing errors during generation.

Relationship modelling remains a persistent challenge, consistently yielding the lowest scores across models and conditions. Future work will therefore focus on relationship-specific mitigation, broader UML diagram types, larger industrial datasets, and human validation of reasoning-level categories. We will also investigate prompt echoing as a potential runtime signal for detecting unsupported structural commitments.

More broadly, this work demonstrates the value of studying LLM hallucinations not only as defects in generated artefacts, but as failures that emerge along a generation process. For software modelling, such a process-oriented view offers a foundation for developing more reliable and controllable LLM-based engineering tools.

\section*{Declarations}

\subsection*{Funding}
The research received no external funding.

\subsection*{Ethical approval}
Not applicable.

\subsection*{Informed consent}
Not applicable.

\subsection*{Author Contributions}
\begin{itemize}
  \item \textbf{Mojdeh Rahmanian}: Conceptualisation, Methodology, Software, Investigation, Data Curation, Formal Analysis, Validation, Visualisation, Writing -- Original Draft.
  \item \textbf{Ashkan Sami}: Conceptualisation, Methodology, Investigation, Data Curation, Validation, Supervision, Writing -- Review \& Editing.
  \item \textbf{Foutse Khomh}: Conceptualisation, Supervision, Writing -- Review \& Editing.
  \item \textbf{Yanchao Yu}: Supervision, Writing -- Review \& Editing.
\end{itemize}

\subsection*{Conflict of Interest}
The authors declare that they have no conflict of interest.

\subsection*{Data Availability}
The reasoning traces, the generated PlantUML code, the prompts, and the evaluation results are available in our GitHub replication package (\url{https://github.com/MojdehRahmanian/HallucinationTaxonomy}).


% \begin{thebibliography}{}
% %
% % and use \bibitem to create references. Consult the Instructions
% % for authors for reference list style.
% % Format for Journal Reference
% Author, Article title, Journal, Volume, page numbers (year)
% Author, Book title, page numbers. Publisher, place (year)
% \end{thebibliography}

\bibliographystyle{spbasic}
\bibliography{references}

@article{li2022competition,
  author       = {Yujia Li and
                  David H. Choi and
                  Junyoung Chung and
                  Nate Kushman and
                  Julian Schrittwieser and
                  R{\'{e}}mi Leblond and
                  Tom Eccles and
                  James Keeling and
                  Felix Gimeno and
                  Agustin Dal Lago and
                  Thomas Hubert and
                  Peter Choy and
                  Cyprien de Masson d'Autume and
                  Igor Babuschkin and
                  Xinyun Chen and
                  Po{-}Sen Huang and
                  Johannes Welbl and
                  Sven Gowal and
                  Alexey Cherepanov and
                  James Molloy and
                  Daniel J. Mankowitz and
                  Esme Sutherland Robson and
                  Pushmeet Kohli and
                  Nando de Freitas and
                  Koray Kavukcuoglu and
                  Oriol Vinyals},
  title        = {Competition-Level Code Generation with AlphaCode},
  journal      = {CoRR},
  volume       = {abs/2203.07814},
  year         = {2022},
  url          = {https://doi.org/10.48550/arXiv.2203.07814},
  doi          = {10.48550/ARXIV.2203.07814},
  eprinttype   = {arXiv},
  eprint       = {2203.07814},
  bibsource    = {dblp computer science bibliography, https://dblp.org}
}

@article{ataei2025elicitron,
  author       = {Mohammadmehdi Ataei and
                  Hyunmin Cheong and
                  Daniele Grandi and
                  Ye Wang and
                  Nigel Morris and
                  Alexander Tessier},
  title        = {Elicitron: {A} Large Language Model Agent-Based Simulation Framework
                  for Design Requirements Elicitation},
  journal      = {J. Comput. Inf. Sci. Eng.},
  volume       = {25},
  number       = {2},
  year         = {2025},
  url          = {https://doi.org/10.1115/1.4067388},
  doi          = {10.1115/1.4067388},
  bibsource    = {dblp computer science bibliography, https://dblp.org}
}

@inproceedings{debari2024evaluating,
  author       = {Daniele De Bari and
                  Giacomo Garaccione and
                  Riccardo Coppola and
                  Marco Torchiano and
                  Luca Ardito},
  editor       = {Xavier Franch and
                  Maya Daneva and
                  Silverio Mart{\'{\i}}nez{-}Fern{\'{a}}ndez and
                  Luigi Quaranta},
  title        = {Evaluating Large Language Models in Exercises of {UML} Class Diagram
                  Modeling},
  booktitle    = {Proceedings of the 18th {ACM/IEEE} International Symposium on Empirical
                  Software Engineering and Measurement, {ESEM} 2024, Barcelona, Spain,
                  October 24-25, 2024},
  pages        = {393--399},
  publisher    = {{ACM}},
  year         = {2024},
  url          = {https://doi.org/10.1145/3674805.3690741},
  doi          = {10.1145/3674805.3690741},
  bibsource    = {dblp computer science bibliography, https://dblp.org}
}

@article{ji2023survey,
  author       = {Ziwei Ji and
                  Nayeon Lee and
                  Rita Frieske and
                  Tiezheng Yu and
                  Dan Su and
                  Yan Xu and
                  Etsuko Ishii and
                  Yejin Bang and
                  Andrea Madotto and
                  Pascale Fung},
  title        = {Survey of Hallucination in Natural Language Generation},
  journal      = {{ACM} Comput. Surv.},
  volume       = {55},
  number       = {12},
  pages        = {248:1--248:38},
  year         = {2023},
  url          = {https://doi.org/10.1145/3571730},
  doi          = {10.1145/3571730},
  bibsource    = {dblp computer science bibliography, https://dblp.org}
}

@article{huang2025survey,
    author       = {Lei Huang and
                  Weijiang Yu and
                  Weitao Ma and
                  Weihong Zhong and
                  Zhangyin Feng and
                  Haotian Wang and
                  Qianglong Chen and
                  Weihua Peng and
                  Xiaocheng Feng and
                  Bing Qin and
                  Ting Liu},
  title        = {A Survey on Hallucination in Large Language Models: Principles, Taxonomy,
                  Challenges, and Open Questions},
  journal      = {{ACM} Trans. Inf. Syst.},
  volume       = {43},
  number       = {2},
  pages        = {42:1--42:55},
  year         = {2025},
  url          = {https://doi.org/10.1145/3703155},
  doi          = {10.1145/3703155},
  bibsource    = {dblp computer science bibliography, https://dblp.org}
}

@inproceedings{raunak2021curious,
  author       = {Vikas Raunak and
                  Arul Menezes and
                  Marcin Junczys{-}Dowmunt},
  editor       = {Kristina Toutanova and
                  Anna Rumshisky and
                  Luke Zettlemoyer and
                  Dilek Hakkani{-}T{\"{u}}r and
                  Iz Beltagy and
                  Steven Bethard and
                  Ryan Cotterell and
                  Tanmoy Chakraborty and
                  Yichao Zhou},
  title        = {The Curious Case of Hallucinations in Neural Machine Translation},
  booktitle    = {Proceedings of the 2021 Conference of the North American Chapter of
                  the Association for Computational Linguistics: Human Language Technologies,
                  {NAACL-HLT} 2021, Online, June 6-11, 2021},
  pages        = {1172--1183},
  publisher    = {Association for Computational Linguistics},
  year         = {2021},
  url          = {https://doi.org/10.18653/v1/2021.naacl-main.92},
  doi          = {10.18653/V1/2021.NAACL-MAIN.92},
  bibsource    = {dblp computer science bibliography, https://dblp.org}
}

@inproceedings{maynez2020faithfulness,
  author       = {Joshua Maynez and
                  Shashi Narayan and
                  Bernd Bohnet and
                  Ryan T. McDonald},
  editor       = {Dan Jurafsky and
                  Joyce Chai and
                  Natalie Schluter and
                  Joel R. Tetreault},
  title        = {On Faithfulness and Factuality in Abstractive Summarization},
  booktitle    = {Proceedings of the 58th Annual Meeting of the Association for Computational
                  Linguistics, {ACL} 2020, Online, July 5-10, 2020},
  pages        = {1906--1919},
  publisher    = {Association for Computational Linguistics},
  year         = {2020},
  url          = {https://doi.org/10.18653/v1/2020.acl-main.173},
  doi          = {10.18653/V1/2020.ACL-MAIN.173},
  bibsource    = {dblp computer science bibliography, https://dblp.org}
}

@article{zhang2025llm,
  author       = {Ziyao Zhang and
                  Chong Wang and
                  Yanlin Wang and
                  Ensheng Shi and
                  Yuchi Ma and
                  Wanjun Zhong and
                  Jiachi Chen and
                  Mingzhi Mao and
                  Zibin Zheng},
  title        = {{LLM} Hallucinations in Practical Code Generation: Phenomena, Mechanism,
                  and Mitigation},
  journal      = {Proc. {ACM} Softw. Eng.},
  volume       = {2},
  number       = {{ISSTA}},
  pages        = {481--503},
  year         = {2025},
  url          = {https://doi.org/10.1145/3728894},
  doi          = {10.1145/3728894},
  bibsource    = {dblp computer science bibliography, https://dblp.org}
}

@inproceedings{belzner2023large,
  author       = {Lenz Belzner and
                  Thomas Gabor and
                  Martin Wirsing},
  editor       = {Bernhard Steffen},
  title        = {Large Language Model Assisted Software Engineering: Prospects, Challenges,
                  and a Case Study},
  booktitle    = {Bridging the Gap Between {AI} and Reality - First International Conference,
                  AISoLA 2023, Crete, Greece, October 23-28, 2023, Proceedings},
  series       = {Lecture Notes in Computer Science},
  volume       = {14380},
  pages        = {355--374},
  publisher    = {Springer},
  year         = {2023},
  url          = {https://doi.org/10.1007/978-3-031-46002-9\_23},
  doi          = {10.1007/978-3-031-46002-9\_23},
  bibsource    = {dblp computer science bibliography, https://dblp.org}
}

@inproceedings{fan2023large,
  author       = {Angela Fan and
                  Beliz Gokkaya and
                  Mark Harman and
                  Mitya Lyubarskiy and
                  Shubho Sengupta and
                  Shin Yoo and
                  Jie M. Zhang},
  title        = {Large Language Models for Software Engineering: Survey and Open Problems},
  booktitle    = {{IEEE/ACM} International Conference on Software Engineering: Future
                  of Software Engineering, ICSE-FoSE 2023, Melbourne, Australia, May
                  14-20, 2023},
  pages        = {31--53},
  publisher    = {{IEEE}},
  year         = {2023},
  url          = {https://doi.org/10.1109/ICSE-FoSE59343.2023.00008},
  doi          = {10.1109/ICSE-FOSE59343.2023.00008},
  bibsource    = {dblp computer science bibliography, https://dblp.org}
}

@article{bozyigit2024generating,
  author       = {Fatma Bozyigit and
                  Tolgahan Bardakci and
                  Alireza Khalilipour and
                  Moharram Challenger and
                  Guus Ramackers and
                  {\"{O}}nder Babur and
                  Michel R. V. Chaudron},
  title        = {Generating domain models from natural language text using {NLP:} a
                  benchmark dataset and experimental comparison of tools},
  journal      = {Softw. Syst. Model.},
  volume       = {23},
  number       = {6},
  pages        = {1493--1511},
  year         = {2024},
  url          = {https://doi.org/10.1007/s10270-024-01176-y},
  doi          = {10.1007/S10270-024-01176-Y},
  bibsource    = {dblp computer science bibliography, https://dblp.org}
}

@inproceedings{zheng2023judging,
  author       = {Lianmin Zheng and
                  Wei{-}Lin Chiang and
                  Ying Sheng and
                  Siyuan Zhuang and
                  Zhanghao Wu and
                  Yonghao Zhuang and
                  Zi Lin and
                  Zhuohan Li and
                  Dacheng Li and
                  Eric P. Xing and
                  Hao Zhang and
                  Joseph E. Gonzalez and
                  Ion Stoica},
  editor       = {Alice Oh and
                  Tristan Naumann and
                  Amir Globerson and
                  Kate Saenko and
                  Moritz Hardt and
                  Sergey Levine},
  title        = {Judging LLM-as-a-Judge with MT-Bench and Chatbot Arena},
  booktitle    = {Advances in Neural Information Processing Systems 36: Annual Conference
                  on Neural Information Processing Systems 2023, NeurIPS 2023, New Orleans,
                  LA, USA, December 10 - 16, 2023},
  year         = {2023},
  url          = {http://papers.nips.cc/paper\_files/paper/2023/hash/91f18a1287b398d378ef22505bf41832-Abstract-Datasets\_and\_Benchmarks.html},
  bibsource    = {dblp computer science bibliography, https://dblp.org}
}

@inproceedings{tong2024codejudge,
  author       = {Weixi Tong and
                  Tianyi Zhang},
  editor       = {Yaser Al{-}Onaizan and
                  Mohit Bansal and
                  Yun{-}Nung Chen},
  title        = {CodeJudge: Evaluating Code Generation with Large Language Models},
  booktitle    = {Proceedings of the 2024 Conference on Empirical Methods in Natural
                  Language Processing, {EMNLP} 2024, Miami, FL, USA, November 12-16,
                  2024},
  pages        = {20032--20051},
  publisher    = {Association for Computational Linguistics},
  year         = {2024},
  url          = {https://doi.org/10.18653/v1/2024.emnlp-main.1118},
  doi          = {10.18653/V1/2024.EMNLP-MAIN.1118},
  bibsource    = {dblp computer science bibliography, https://dblp.org}
}

@article{gu2025survey,
  author       = {Jiawei Gu and
                  Xuhui Jiang and
                  Zhichao Shi and
                  Hexiang Tan and
                  Xuehao Zhai and
                  Chengjin Xu and
                  Wei Li and
                  Yinghan Shen and
                  Shengjie Ma and
                  Honghao Liu and
                  Yuanzhuo Wang and
                  Jian Guo},
  title        = {A Survey on LLM-as-a-Judge},
  journal      = {CoRR},
  volume       = {abs/2411.15594},
  year         = {2024},
  url          = {https://doi.org/10.48550/arXiv.2411.15594},
  doi          = {10.48550/ARXIV.2411.15594},
  eprinttype   = {arXiv},
  eprint       = {2411.15594},
  bibsource    = {dblp computer science bibliography, https://dblp.org}
}

@article{marjanovic2025deepseek,
  author       = {Sara Vera Marjanovic and
                  Arkil Patel and
                  Vaibhav Adlakha and
                  Milad Aghajohari and
                  Parishad BehnamGhader and
                  Mehar Bhatia and
                  Aditi Khandelwal and
                  Austin Kraft and
                  Benno Krojer and
                  Xing Han L{\`{u}} and
                  Nicholas Meade and
                  Dongchan Shin and
                  Amirhossein Kazemnejad and
                  Gaurav Kamath and
                  Marius Mosbach and
                  Karolina Stanczak and
                  Siva Reddy},
  title        = {DeepSeek-R1 Thoughtology: Let's think about {LLM} reasoning},
  journal      = {Trans. Mach. Learn. Res.},
  volume       = {2026},
  year         = {2026},
  url          = {https://openreview.net/forum?id=BZwKsiRnJI},
  bibsource    = {dblp computer science bibliography, https://dblp.org}
}

@article{hou2024large,
   author       = {Xinyi Hou and
                  Yanjie Zhao and
                  Yue Liu and
                  Zhou Yang and
                  Kailong Wang and
                  Li Li and
                  Xiapu Luo and
                  David Lo and
                  John Grundy and
                  Haoyu Wang},
  title        = {Large Language Models for Software Engineering: {A} Systematic Literature
                  Review},
  journal      = {{ACM} Trans. Softw. Eng. Methodol.},
  volume       = {33},
  number       = {8},
  pages        = {220:1--220:79},
  year         = {2024},
  url          = {https://doi.org/10.1145/3695988},
  doi          = {10.1145/3695988},
  bibsource    = {dblp computer science bibliography, https://dblp.org}
}

@article{camara2024systematic,
  author       = {Javier C{\'{a}}mara and
                  Javier Troya and
                  Lola Burgue{\~{n}}o and
                  Antonio Vallecillo},
  title        = {On the assessment of generative {AI} in modeling tasks: an experience
                  report with ChatGPT and {UML}},
  journal      = {Softw. Syst. Model.},
  volume       = {22},
  number       = {3},
  pages        = {781--793},
  year         = {2023},
  url          = {https://doi.org/10.1007/s10270-023-01105-5},
  doi          = {10.1007/S10270-023-01105-5},
  bibsource    = {dblp computer science bibliography, https://dblp.org}
}

@inproceedings{ferrari2024model,
  author       = {Alessio Ferrari and
                  Sallam Abualhaija and
                  Chetan Arora},
  title        = {Model Generation with LLMs: From Requirements to {UML} Sequence Diagrams},
  booktitle    = {32nd {IEEE} International Requirements Engineering Conference, {RE}
                  2024 - Workshops, Reykjavik, Iceland, June 24-25, 2024},
  pages        = {291--300},
  publisher    = {{IEEE}},
  year         = {2024},
  url          = {https://doi.org/10.1109/REW61692.2024.00044},
  doi          = {10.1109/REW61692.2024.00044},
  bibsource    = {dblp computer science bibliography, https://dblp.org}
}

@article{freeman2026epistemic,
  author       = {Brian S. Freeman and
                  Adam Kicklighter and
                  Matt Erdman and
                  Zach Gordon},
  title        = {Toward Epistemic Stability: Engineering Consistent Procedures for
                  Industrial {LLM} Hallucination Reduction},
  journal      = {CoRR},
  volume       = {abs/2603.10047},
  year         = {2026},
  url          = {https://doi.org/10.48550/arXiv.2603.10047},
  doi          = {10.48550/ARXIV.2603.10047},
  eprinttype   = {arXiv},
  eprint       = {2603.10047},
  bibsource    = {dblp computer science bibliography, https://dblp.org}
}

@article{zhang2025mitigating,
   author  = {Zhang, Hongjie and Deng, Hourui and Ou, Jie and Feng, Chaosheng},
   title   = {Mitigating spatial hallucination in large language models for path planning via prompt engineering},
   journal = {Scientific Reports},
   volume  = {15},
   pages   = {8903},
   year    = {2025},
   doi     = {10.1038/s41598-025-93601-5}
 }

@inproceedings{conrardy2024image,
  author       = {Aaron David Conrardy and
                  Jordi Cabot},
  editor       = {Hessa Alfraihi and
                  Francesco Basciani and
                  Georgiana Caltais and
                  Nicolas Ferry and
                  Jos{\'{e}} Antonio Hern{\'{a}}ndez L{\'{o}}pez and
                  Ludovico Iovino and
                  Robbert Jongeling and
                  Stefan Klikovits and
                  Shekoufeh Kolahdouz Rahimi and
                  Riccardo Rubei and
                  Sobhan Yassipour Tehrani and
                  Javier Troya and
                  Mairieli Wessel and
                  Vadim Zaytsev},
  title        = {From image to {UML:} First results of image-based {UML} diagram generation
                  using LLMs},
  booktitle    = {Proceedings of the {STAF} 2024 Workshops: AgileMDE 2024, {LLM4MDE}
                  2024, and MeSS 2024 co-located with the International Conference on
                  Software Technologies: Applications and Foundations {(STAF} 2024)
                  Enschede, The Netherlands, July 8-11, 2024},
  series       = {{CEUR} Workshop Proceedings},
  volume       = {3727},
  pages        = {55--65},
  publisher    = {CEUR-WS.org},
  year         = {2024},
  url          = {https://ceur-ws.org/Vol-3727/paper5.pdf},
  bibsource    = {dblp computer science bibliography, https://dblp.org}
}

@article{shim2025cpr,
  author       = {Jung{-}Woo Shim and
                  Yeong{-}Joon Ju and
                  Ji{-}Hoon Park and
                  Seong{-}Whan Lee},
  title        = {{CPR:} Mitigating Large Language Model Hallucinations with Curative
                  Prompt Refinement},
  journal      = {CoRR},
  volume       = {abs/2510.12029},
  year         = {2025},
  url          = {https://doi.org/10.48550/arXiv.2510.12029},
  doi          = {10.48550/ARXIV.2510.12029},
  eprinttype   = {arXiv},
  eprint       = {2510.12029},
  bibsource    = {dblp computer science bibliography, https://dblp.org}
}

@inproceedings{verbruggen2025golden,
  author       = {Charlotte Verbruggen and
                  Lukas Netz and
                  Philipp{-}Lorenz Glaser and
                  Marion Scholz and
                  Christian Huemer and
                  Marco Calamo and
                  Bernhard Rumpe and
                  Monique Snoeck and
                  Dominik Bork},
  title        = {Toward a Community-Curated Golden Dataset of {UML} Models},
  booktitle    = {28th {ACM/IEEE} International Conference on Model Driven Engineering
                  Languages and Systems, {MODELS} 2025 - Companion, Grand Rapids, MI,
                  USA, October 5-10, 2025},
  pages        = {43--50},
  publisher    = {{IEEE}},
  year         = {2025},
  url          = {https://doi.org/10.1109/MODELS-C68889.2025.00012},
  doi          = {10.1109/MODELS-C68889.2025.00012},
  bibsource    = {dblp computer science bibliography, https://dblp.org}
}

@article{lindland1994understanding,
  author       = {Odd Ivar Lindland and
                  Guttorm Sindre and
                  Arne S{\o}lvberg},
  title        = {Understanding Quality in Conceptual Modeling},
  journal      = {{IEEE} Softw.},
  volume       = {11},
  number       = {2},
  pages        = {42--49},
  year         = {1994},
  url          = {https://doi.org/10.1109/52.268955},
  doi          = {10.1109/52.268955},
  bibsource    = {dblp computer science bibliography, https://dblp.org}
}

@article{chillarege1992orthogonal,
  author       = {Ram Chillarege and
                  Inderpal S. Bhandari and
                  Jarir K. Chaar and
                  Michael J. Halliday and
                  Diane S. Moebus and
                  Bonnie K. Ray and
                  Man{-}Yuen Wong},
  title        = {Orthogonal Defect Classification - {A} Concept for In-Process Measurements},
  journal      = {{IEEE} Trans. Software Eng.},
  volume       = {18},
  number       = {11},
  pages        = {943--956},
  year         = {1992},
  url          = {https://doi.org/10.1109/32.177364},
  doi          = {10.1109/32.177364},
  bibsource    = {dblp computer science bibliography, https://dblp.org}
}

@inproceedings{humbatova2020taxonomy,
  author       = {Nargiz Humbatova and
                  Gunel Jahangirova and
                  Gabriele Bavota and
                  Vincenzo Riccio and
                  Andrea Stocco and
                  Paolo Tonella},
  editor       = {Gregg Rothermel and
                  Doo{-}Hwan Bae},
  title        = {Taxonomy of real faults in deep learning systems},
  booktitle    = {{ICSE} '20: 42nd International Conference on Software Engineering,
                  Seoul, South Korea, 27 June - 19 July, 2020},
  pages        = {1110--1121},
  publisher    = {{ACM}},
  year         = {2020},
  url          = {https://doi.org/10.1145/3377811.3380395},
  doi          = {10.1145/3377811.3380395},
  bibsource    = {dblp computer science bibliography, https://dblp.org}
}

@article{liu2024exploring,
  author       = {Fang Liu and
                  Yang Liu and
                  Lin Shi and
                  Houkun Huang and
                  Ruifeng Wang and
                  Zhen Yang and
                  Li Zhang},
  title        = {Exploring and Evaluating Hallucinations in LLM-Powered Code Generation},
  journal      = {CoRR},
  volume       = {abs/2404.00971},
  year         = {2024},
  url          = {https://doi.org/10.48550/arXiv.2404.00971},
  doi          = {10.48550/ARXIV.2404.00971},
  eprinttype   = {arXiv},
  eprint       = {2404.00971},
  bibsource    = {dblp computer science bibliography, https://dblp.org}
}

@inproceedings{pan2024lost,
  author       = {Rangeet Pan and
                  Ali Reza Ibrahimzada and
                  Rahul Krishna and
                  Divya Sankar and
                  Lambert Pouguem Wassi and
                  Michele Merler and
                  Boris Sobolev and
                  Raju Pavuluri and
                  Saurabh Sinha and
                  Reyhaneh Jabbarvand},
  title        = {Lost in Translation: {A} Study of Bugs Introduced by Large Language
                  Models while Translating Code},
  booktitle    = {Proceedings of the 46th {IEEE/ACM} International Conference on Software
                  Engineering, {ICSE} 2024, Lisbon, Portugal, April 14-20, 2024},
  pages        = {82:1--82:13},
  publisher    = {{ACM}},
  year         = {2024},
  url          = {https://doi.org/10.1145/3597503.3639226},
  doi          = {10.1145/3597503.3639226},
  bibsource    = {dblp computer science bibliography, https://dblp.org}
}

@article{lanham2023measuring,
  author       = {Tamera Lanham and
                  Anna Chen and
                  Ansh Radhakrishnan and
                  Benoit Steiner and
                  Carson Denison and
                  Danny Hernandez and
                  Dustin Li and
                  Esin Durmus and
                  Evan Hubinger and
                  Jackson Kernion and
                  Kamile Lukosiute and
                  Karina Nguyen and
                  Newton Cheng and
                  Nicholas Joseph and
                  Nicholas Schiefer and
                  Oliver Rausch and
                  Robin Larson and
                  Sam McCandlish and
                  Sandipan Kundu and
                  Saurav Kadavath and
                  Shannon Yang and
                  Thomas Henighan and
                  Timothy Maxwell and
                  Timothy Telleen{-}Lawton and
                  Tristan Hume and
                  Zac Hatfield{-}Dodds and
                  Jared Kaplan and
                  Jan Brauner and
                  Samuel R. Bowman and
                  Ethan Perez},
  title        = {Measuring Faithfulness in Chain-of-Thought Reasoning},
  journal      = {CoRR},
  volume       = {abs/2307.13702},
  year         = {2023},
  url          = {https://doi.org/10.48550/arXiv.2307.13702},
  doi          = {10.48550/ARXIV.2307.13702},
  eprinttype   = {arXiv},
  eprint       = {2307.13702},
  bibsource    = {dblp computer science bibliography, https://dblp.org}
}

@inproceedings{turpin2023language,
  author       = {Miles Turpin and
                  Julian Michael and
                  Ethan Perez and
                  Samuel R. Bowman},
  editor       = {Alice Oh and
                  Tristan Naumann and
                  Amir Globerson and
                  Kate Saenko and
                  Moritz Hardt and
                  Sergey Levine},
  title        = {Language Models Don't Always Say What They Think: Unfaithful Explanations
                  in Chain-of-Thought Prompting},
  booktitle    = {Advances in Neural Information Processing Systems 36: Annual Conference
                  on Neural Information Processing Systems 2023, NeurIPS 2023, New Orleans,
                  LA, USA, December 10 - 16, 2023},
  year         = {2023},
  url          = {http://papers.nips.cc/paper\_files/paper/2023/hash/ed3fea9033a80fea1376299fa7863f4a-Abstract-Conference.html},
  bibsource    = {dblp computer science bibliography, https://dblp.org}
}

@article{sun_detection_2025,
  author       = {Zhongxiang Sun and
                  Qipeng Wang and
                  Haoyu Wang and
                  Xiao Zhang and
                  Jun Xu},
  title        = {Detection and Mitigation of Hallucination in Large Reasoning Models:
                  {A} Mechanistic Perspective},
  journal      = {CoRR},
  volume       = {abs/2505.12886},
  year         = {2025},
  url          = {https://doi.org/10.48550/arXiv.2505.12886},
  doi          = {10.48550/ARXIV.2505.12886},
  eprinttype   = {arXiv},
  eprint       = {2505.12886},
  bibsource    = {dblp computer science bibliography, https://dblp.org}
}

@inproceedings{lee_evaluating_2025,
 author       = {Jinu Lee and
                  Julia Hockenmaier},
  editor       = {Christos Christodoulopoulos and
                  Tanmoy Chakraborty and
                  Carolyn Rose and
                  Violet Peng},
  title        = {Evaluating Step-by-step Reasoning Traces: {A} Survey},
  booktitle    = {Findings of the Association for Computational Linguistics: {EMNLP}
                  2025, Suzhou, China, November 4-9, 2025},
  pages        = {1789--1814},
  publisher    = {Association for Computational Linguistics},
  year         = {2025},
  url          = {https://doi.org/10.18653/v1/2025.findings-emnlp.94},
  doi          = {10.18653/V1/2025.FINDINGS-EMNLP.94},
  bibsource    = {dblp computer science bibliography, https://dblp.org}
}

@article{hripcsak2005agreement,
  author  = {Hripcsak, George and Rothschild, Adam S.},
  title   = {Agreement, the F-Measure, and Reliability in Information Retrieval},
  journal = {Journal of the American Medical Informatics Association},
  volume  = {12},
  number  = {3},
  pages   = {296--298},
  year    = {2005},
  doi     = {10.1197/jamia.M1733}
}

@inproceedings{lewis2020retrieval,
  author       = {Patrick Lewis and
                  Ethan Perez and
                  Aleksandra Piktus and
                  Fabio Petroni and
                  Vladimir Karpukhin and
                  Naman Goyal and
                  Heinrich K{\"{u}}ttler and
                  Mike Lewis and
                  Wen{-}tau Yih and
                  Tim Rockt{\"{a}}schel and
                  Sebastian Riedel and
                  Douwe Kiela},
  editor       = {Hugo Larochelle and
                  Marc'Aurelio Ranzato and
                  Raia Hadsell and
                  Maria{-}Florina Balcan and
                  Hsuan{-}Tien Lin},
  title        = {Retrieval-Augmented Generation for Knowledge-Intensive {NLP} Tasks},
  booktitle    = {Advances in Neural Information Processing Systems 33: Annual Conference
                  on Neural Information Processing Systems 2020, NeurIPS 2020, December
                  6-12, 2020, virtual},
  year         = {2020},
  url          = {https://proceedings.neurips.cc/paper/2020/hash/6b493230205f780e1bc26945df7481e5-Abstract.html},
  bibsource    = {dblp computer science bibliography, https://dblp.org}
}

@inproceedings{shuster2021retrieval,
  author       = {Kurt Shuster and
                  Spencer Poff and
                  Moya Chen and
                  Douwe Kiela and
                  Jason Weston},
  editor       = {Marie{-}Francine Moens and
                  Xuanjing Huang and
                  Lucia Specia and
                  Scott Wen{-}tau Yih},
  title        = {Retrieval Augmentation Reduces Hallucination in Conversation},
  booktitle    = {Findings of the Association for Computational Linguistics: {EMNLP}
                  2021, Virtual Event / Punta Cana, Dominican Republic, 16-20 November,
                  2021},
  series       = {Findings of {ACL}},
  volume       = {{EMNLP} 2021},
  pages        = {3784--3803},
  publisher    = {Association for Computational Linguistics},
  year         = {2021},
  url          = {https://doi.org/10.18653/v1/2021.findings-emnlp.320},
  doi          = {10.18653/V1/2021.FINDINGS-EMNLP.320},
  bibsource    = {dblp computer science bibliography, https://dblp.org}
}

@inproceedings{dhuliawala2024chain,
  author       = {Shehzaad Dhuliawala and
                  Mojtaba Komeili and
                  Jing Xu and
                  Roberta Raileanu and
                  Xian Li and
                  Asli Celikyilmaz and
                  Jason Weston},
  editor       = {Lun{-}Wei Ku and
                  Andre Martins and
                  Vivek Srikumar},
  title        = {Chain-of-Verification Reduces Hallucination in Large Language Models},
  booktitle    = {Findings of the Association for Computational Linguistics, {ACL} 2024,
                  Bangkok, Thailand and virtual meeting, August 11-16, 2024},
  series       = {Findings of {ACL}},
  volume       = {{ACL} 2024},
  pages        = {3563--3578},
  publisher    = {Association for Computational Linguistics},
  year         = {2024},
  url          = {https://doi.org/10.18653/v1/2024.findings-acl.212},
  doi          = {10.18653/V1/2024.FINDINGS-ACL.212},
  bibsource    = {dblp computer science bibliography, https://dblp.org}
}

@inproceedings{manakul2023selfcheckgpt,
  author       = {Potsawee Manakul and
                  Adian Liusie and
                  Mark J. F. Gales},
  editor       = {Houda Bouamor and
                  Juan Pino and
                  Kalika Bali},
  title        = {SelfCheckGPT: Zero-Resource Black-Box Hallucination Detection for
                  Generative Large Language Models},
  booktitle    = {Proceedings of the 2023 Conference on Empirical Methods in Natural
                  Language Processing, {EMNLP} 2023, Singapore, December 6-10, 2023},
  pages        = {9004--9017},
  publisher    = {Association for Computational Linguistics},
  year         = {2023},
  url          = {https://doi.org/10.18653/v1/2023.emnlp-main.557},
  doi          = {10.18653/V1/2023.EMNLP-MAIN.557},
  bibsource    = {dblp computer science bibliography, https://dblp.org}
}

@inproceedings{chen2023automated,
  author       = {Kua Chen and
                  Yujing Yang and
                  Boqi Chen and
                  Jos{\'{e}} Antonio Hern{\'{a}}ndez L{\'{o}}pez and
                  Gunter Mussbacher and
                  D{\'{a}}niel Varr{\'{o}}},
  title        = {Automated Domain Modeling with Large Language Models: {A} Comparative
                  Study},
  booktitle    = {26th {ACM/IEEE} International Conference on Model Driven Engineering
                  Languages and Systems, {MODELS} 2023, V{\"{a}}ster{\aa}s, Sweden,
                  October 1-6, 2023},
  pages        = {162--172},
  publisher    = {{IEEE}},
  year         = {2023},
  url          = {https://doi.org/10.1109/MODELS58315.2023.00037},
  doi          = {10.1109/MODELS58315.2023.00037},
  bibsource    = {dblp computer science bibliography, https://dblp.org}
}

\end{document}